\documentclass[a4paper,10pt,twocolumn]{article}

\usepackage[hidelinks]{hyperref}

\usepackage{graphicx} 

\usepackage{amsmath}
\def\be{\begin{equation}}
\def\ee{\end{equation}}
\def\bea{\begin{eqnarray}}
\def\eea{\end{eqnarray}}

\usepackage[T1]{fontenc}
\usepackage{lmodern}
\usepackage[english]{babel}
\usepackage{microtype}
\usepackage[medium,compact]{titlesec}

\usepackage{gensymb}
\usepackage{textcomp}

\usepackage[a4paper,left=2cm,right=2cm,top=2cm,bottom=2cm]{geometry}

\usepackage[version=4]{mhchem}
\usepackage{chemmacros}

\usepackage[separate-uncertainty=true]{siunitx} 
\DeclareSIUnit\Molar{\textsc{m}}

\usepackage{color}
\definecolor{red}{rgb}{0.75,0,0}
\definecolor{blue}{rgb}{0,0,0.75}
\definecolor{green}{rgb}{0,0.5,0}

\usepackage[square,comma,numbers,compress]{natbib}

\usepackage{tabularx}
\usepackage{booktabs}
\AtBeginDocument{
\heavyrulewidth=.08em
\lightrulewidth=.05em
\cmidrulewidth=.03em
\belowrulesep=.4ex
\belowbottomsep=0pt
\aboverulesep=.4ex
\abovetopsep=0pt
\cmidrulesep=\doublerulesep
\cmidrulekern=.5em
\defaultaddspace=.5em
}

\usepackage{authblk}

\title{{\Large\bf Colloidal supported lipid bilayers for self-assembly}}
\author[$\ast$,1,2]{Melissa~Rinaldin}
\author[$\ast$,1]{Ruben~W.~Verweij}
\author[3]{Indrani~Chakraborty}
\author[1]{Daniela~J.~Kraft}
\affil[$\ast$]{These authors contributed equally to the work.}
\affil[1]{Huygens-Kamerlingh Onnes Lab, Universiteit Leiden, P. O. Box 9504, 2300 RA Leiden, The Netherlands}
\affil[2]{Instituut-Lorentz, Universiteit Leiden, P.O. Box 9506, 2300 RA Leiden, The Netherlands}
\affil[3]{School of Chemistry, Raymond and Beverly Sackler Faculty of Exact Sciences, Tel Aviv University, Tel Aviv 69978, Israel}
\affil[ ]{Corresponding email: \url{kraft@physics.leidenuniv.nl}}

\date{}

\begin{document}

\twocolumn[{%
  \begin{@twocolumnfalse}
    \maketitle
    \begin{abstract} 

        \bfseries\noindent The use of colloidal supported lipid bilayers
        (CSLBs) has recently been extended to create colloidal joints, that -  in analogy to their macroscopic counterparts - can flexibly connect colloidal particles. These novel
        elements enable the assembly of structures with internal degrees of flexibility, but rely on
        previously unappreciated properties: the simultaneous fluidity of the
        bilayer, lateral mobility of inserted (linker) molecules and colloidal
        stability. Here we characterize every step in the manufacturing of
        CSLBs in view of these requirements using confocal microscopy and fluorescence recovery after photobleaching (FRAP). Specifically, we have studied the
        influence of different particle properties (roughness, surface charge,
        chemical composition, polymer coating) on the quality and mobility of
        the supported bilayer. We find that the insertion of lipopolymers in
        the bilayer can affect its homogeneity and fluidity. We improve the
        colloidal stability by inserting lipopolymers or double-stranded inert
        DNA into the bilayer.  Finally, we include surface-mobile DNA linkers
        and use FRAP to characterize their lateral mobility both in their
        freely diffusive and bonded state. Our work offers a collection of
        experimental tools for working with CSLBs in applications ranging from
        controlled bottom-up self-assembly to model membrane
        studies.\vspace{1em}

    \end{abstract} 
  \end{@twocolumnfalse}
}]

\section*{Introduction} 

Colloidal supported lipid bilayers (CSLBs) are used in a diverse range of
research areas and applications \cite{Troutier2007}, from drug delivery
\cite{Carmona-Ribeiro2012, Li2014, Savarala2010}, bio-sensing
\cite{Castellana2006, Chemburu2010}, membrane manipulation \cite{Brouwer2015}
and cell biology \cite{Sackmann2007a, Madwar2015, Mashaghi2013} to fundamental
studies on lipid phase separation \cite{Rinaldin2018, Fonda2018} and self-assembly
\cite{VanDerMeulen2013, VanDerMeulen2014, VanderMeulen2015, Chakraborty2016a}.
The presence of a lipid bilayer around nano- or micrometer-sized solid
particles or droplets provides biomimetic properties and a platform for further
functionalization. One intriguing recent example used DNA-based linkers to functionalize the lipid bilayer thereby enabling self-assembly of the underlying colloidal particles or droplets into flexible structures.
\cite{VanDerMeulen2013,Feng2013a, VanDerMeulen2014, VanderMeulen2015,
Chakraborty2016a,McMullen2018}. Within such a structure, the colloidal elements
can move over each other's surface while remaining strongly and specifically
bonded. This new type of bonding enables fundamental studies on structures with
internal degrees of flexibility, such as the self-assembly of novel crystal
phases and their phase transitions \cite{Kohlstedt2013, Ortiz2014, Smallenburg,
Hu2018}. Furthermore, these complex colloids have great potential for smart
drug delivery techniques \cite{Carmona-Ribeiro2012}, photonic band-gap
materials \cite{Joannopoulos1997,Lin2005} and wet computing
\cite{Phillips2014a}.

CSLBs are particularly suitable and versatile building blocks for the assembly
of floppy structures, because they combine the best qualities of free standing
bilayers (vesicles) and colloids. Vesicles, upon applications of linkers
\cite{Hadorn2010}, can connect into flexible structures, but are unstable to
small disturbances, heterogenous in size and easily deformable. Colloidal
particles are available in diverse materials and with a variety of stable
shapes, and can be assembled after functionalization with surface-bound DNA
linkers.\cite{Wang2012,Wang2015a}. However, the obtained structures are often
rigid due to the immobility of the linking groups on the particles' surface and
are non-equilibrium structures due to a ``hit-and-stick'' aggregation process
\cite{Schade2013}. Emulsions coated with lipid monolayers and DNA-linkers that
are mobile on the droplet interface posses both interaction specificity and
bond flexibility \cite{Feng2013a, Zhang2017, McMullen2018}. Therefore, they
assemble into flexible structures in a controlled fashion, but their shape is
limited to spheres and they deform upon binding. Conversely, CSLBs consist of
colloidal particles which provide a stable support for the lipid bilayer that
is tunable in shape, size and material. The range of shapes for colloidal
particles comprises, among others, spheres, cubes, rods, and (a)symmetric
dumbbell particles, and their sizes range from hundreds of nanometers to several
micrometers. They can be produced reliably with a narrow size distribution and
are commercially available. Additionally, CLSBs feature a lipid bilayer on the
surface of the colloids which creates a liquid film for molecules, such as DNA
linkers, to freely move in. This allows for binding particles specifically, and
yet non-rigidly, making the assembly of floppy structures possible
\cite{VanDerMeulen2013, VanDerMeulen2014, VanderMeulen2015, Chakraborty2016a,
Zhang2017}. 

To obtain flexible instead of rigid structures, it is vital that the linker
molecules which are inserted into the lipid bilayer are free to move over the
surface of the CSLBs. Their lateral mobility relies on the fluidity and
homogeneity of the bilayer, which in turn depend on the linker concentration
\cite{Chakraborty2016a} and lipid composition. The lipids need to be in the
fluid state under experimental conditions, and this may be impeded by
bilayer-surface interactions. Similarly, the success of experiments studying
the phase separation of lipid bilayers on anisotropic colloidal supports relies
on the fluidity and homogeneity of the bilayer \cite{Rinaldin2018, Fonda2018}. Finally,
controlling the self-assembly pathway through complementary DNA linkers implies
that all other non-specific interactions need to be suppressed. In other words,
CSLBs need to have sufficient colloidal stability. To the best of our
knowledge, these requirements of membrane homogeneity and fluidity plus
colloidal stability have not been studied simultaneously. However, they are of
key importance for using CSLBs in self-assembly and model membrane studies,
while possibly having wider implications for all other applications. 

Here, we carefully characterize every stage in the preparation of CSLBs
specifically related to these three properties. First, we study the effect of
the material properties of the particles and the use of polymers on the
membrane fluidity and homogeneity. Then, we investigate the influence of
lipopolymers and inert double-stranded DNA on the colloidal stability of the
CSLBs. Subsequently, we include DNA-based linkers connected to hydrophobic
anchors and characterize their diffusion in the bilayer. Finally, we show that
when using the optimal experimental parameters determined by this study, CSLBs
self-assemble into flexibly linked structures that are freely-jointed.

\section*{Experimental Section}

\subsection*{Reagents}

\subsubsection*{Chemicals}

\iupac{1-pal|mit|oyl-2-o|le|oyl-sn-gly|ce|ro-3-phos|pho|cho|line} (POPC),
\iupac{L-$\alpha$-Phos|pha|ti|dyl|e|tha|nol|a|mine-N-(DOPE liss|a|mine
rhod|a|mine B sul|fo|nyl)}, \iupac{23-(di|pyr|ro|meth|ene|bo|ron
di|flu|o|ride)-24-nor|cho|les|te|rol} (TopFluor\textregistered-Cholesterol),
\iupac{($\Delta$9-Cis) 1,2-di|o|le|oyl-sn-gly|ce|ro-3-phos|pho|cho|line}
(DOPC),
\iupac{1,2-di|o|le|oyl-sn-gly|ce|ro-3-phos|pho|e|tha|nol|a|mine-N-[me|tho|xy|(po|ly|e|thy|lene
gly|col)|-2000]} (DOPE-PEG(2000)), DOPE-PEG(3000) and DOPE-PEG(5000) were
purchased from Avanti Polar Lipids.
\iupac{4-(2-hy|dro|xy|e|thyl)-1-pi|per|a|zi|ne|e|tha|ne|sul|fo|nic ac|id}
(HEPES, $\geq$\SI{99.5}{\percent}) and \iupac{cal|ci|um chlo|ride} (\ce{CaCl2},
\iupac{Cal|ci|um|chlo|rid Di|hy|drat}, $\geq$\SI{99}{\percent}) were purchased
from Carl Roth. \iupac{So|di|um chlo|ride} (\ce{NaCl}, extra pure),
\iupac{hy|dro|gen per|o|xide} (\ce{H2O2}, \SI{35}{\percent w/w}),
\iupac{a|cryl|a|mide} (\SI{98.5}{\percent}, extra pure),
\iupac{N,|N,|N$'$,|N$'$-te|tra|me|thyl|e|thyl|ene|di|a|mine} (TEMED,
\SI{99}{\percent}), \iupac{am|mo|ni|um per|sul|fate} (APS, \SI{98}{\percent}),
\iupac{so|di|um hy|dro|xi|de} (\ce{NaOH}, \SI{98.5}{\percent}) and
\iupac{so|di|um a|zi|de} (\ce{NaN3}, \SI{99}{\percent}, extra pure) were
purchased from Acros Organics. \iupac{Hell|ma|nex}\texttrademark~III,
\iupac{am|mo|ni|um hy|drox|ide} (\ce{NH4OH}, 28-30~\si{\percent w/w}),
\iupac{3-(tri|me|tho|xy|si|lyl)|pro|pyl me|tha|cry|late} (TPM,
\SI{98}{\percent}), Pluronic\textregistered~F-127, \iupac{di|po|tas|si|um
phos|phate} (\ce{K2HPO4}, $\geq$\SI{99}{\percent}), ethanol
($\geq$\SI{99.8}{\percent}), sodium dodecyl sulfate (SDS,
$\geq$\SI{98.5}{\percent}), polyvinylpyrrolidone (PVP, average Mw \num{40000}),
itaconic acid ($\geq$\SI{99}{\percent}),
\iupac{3-a|mi|no|pro|pyl|tri|e|tho|xy|si|lane} (\SI{99}{\percent}) and acetic
acid (\SI{99.8}{\percent}) were purchased from Sigma-Aldrich.  Magnesium
chloride (\ce{MgCl2}, for analysis) was purchased from Merck. All solutions
were prepared with Milli-Q water (Milli-Q Gradient A10).

\subsubsection*{Buffers}

HEPES buffer type 1 was made with \SI{115}{\milli\Molar} \ce{NaCl},
\SI{1.2}{\milli\Molar} \ce{CaCl2}, \SI{1.2}{\milli\Molar} \ce{MgCl2},
\SI{2.4}{\milli\Molar} \ce{K2HPO4} and \SI{20}{\milli\Molar} HEPES. HEPES
buffer type 2 consisted of \SI{10}{\milli\Molar} HEPES, \SI{40}{\milli\Molar}
\ce{NaCl}, \SI{2}{\milli\Molar} \ce{CaCl2} and \SI{3}{\milli\Molar} \ce{NaN3}.
HEPES buffer type 3 consisted of \SI{10}{\milli\Molar} HEPES,
\SI{40}{\milli\Molar} \ce{NaCl} and \SI{3}{\milli\Molar} \ce{NaN3}. The
buffers were prepared by mixing all reagents in the appropriate amounts in
fresh Milli-Q water. After mixing, the pH was adjusted to 7.4 using \ce{NaOH}. 

\subsubsection*{Particles} 

Commercial silica spheres
(\SIlist{2.06\pm0.05;2.12\pm0.06;7\pm0.29}{\micro\meter}) were synthesized by
Microparticles GmbH, using a St\"{o}ber method where tetraethoxysilane (TEOS)
reacts with water and bases in an ethanolic solution (sol-gel process).
Commercial polystyrene particles (\SIlist{2\pm0.03}{\micro\meter}) were
obtained from Sigma Aldrich. Hematite cubic particles
(\SIlist{1.76\pm0.08}{\micro\meter}) were made following the protocol of
Sugimoto et al. \cite{Sugimoto1992} and coated according to Rossi et al.
\cite{Rossi2011}. Polystyrene-3-(Trimethoxysilyl)propyl methacrylate
(Polystyrene-TPM) particles (\SIlist{1.24\pm0.04}{\micro\meter}) with varying
asperity were synthesized and coated with silica following the protocol of
Meester et al. \cite{Meester2016}. TPM particles
(\SIlist{2.03\pm0.06}{\micro\meter}) were made following the protocol of Van
der Wel et al. \cite{VanDerWel2017TPM}. TPM particles functionalized with
carboxyl groups (\SIlist{2.71\pm0.14}{\micro\meter}), or amino groups
(\SIlist{2.14\pm0.07}{\micro\meter}) were prepared by synthesizing according to
\cite{VanDerWel2017TPM} and then functionalizing according to
\cite{DohertyTBP}. Briefly, amine or carboxylic acid groups were incorporated
onto the TPM surface by addition of either 3-aminopropyltriethoxysilane or
itaconic acid, respectively, during the emulsification stage. Polystyrene
particles with carboxyl groups (\SIlist{1.99\pm0.15}{\micro\meter}) were
synthesized according to Appel et al. \cite{Appel2013}. Polystyrene-TPM
particles of spherical, symmetric and asymmetric dumbbell shape were made and
coated with silica following the protocols reported in Rinaldin et al.
\cite{Rinaldin2018}.

\subsubsection*{DNA oligonucleotides}

All DNA strands were synthesized as single stranded DNA, purified using reverse
phase high-performance liquid chromatography (HPLC-RP) and checked using
matrix-assisted laser desorption/ionization time-of-flight mass spectrometry
(Maldi-TOF MS) by Kaneka Eurogentec S.A. We used double-stranded inert DNA for
steric stabilization and double-stranded DNA linkers with a sticky end for
binding. Both types of DNA have a hydrophobic anchor (double Stearyl/double
cholesterol for linker and double cholesterol for inert DNA) connected to a
short carbon chain which is then connected to the oligonucleotide. The linkers
are additionally functionalized with a fluorescent dye. All strands, including
all functionalizations, are listed in Table S1 of the Supporting Information.
These DNA strands were hybridized by mixing the single strands in a 1:1 molar
ratio in HEPES buffer type 3. The solution was then heated in an oven to
\SI{90}{\degreeCelsius} for \SI{30}{\minute}, after which the oven was turned
off and the solution was allowed to cool down slowly overnight in the closed
oven. Strand 1 and 2 are hybridized to form ``\SI{10}{\nm} Inert DNA'', 3 and 4
are hybridized to form ``\SI{20}{\nm} Inert DNA'', 4 and 5 form ``\SI{20}{\nm}
Linker A'', 4 and 6 make ``\SI{20}{\nm} Linker A$'$'', 7 and 8 form
``\SI{30}{\nm} Linker A'' and, finally, 7 and 9 are hybridized to form
``\SI{30}{\nm} Linker A$'$''. The linkers ``A'' have a single-stranded sticky
end (indicated by cursive text in Table S1 of the Supporting Information)
that is complementary to the single-stranded end of linkers ``A$'$''.

\subsection*{Preparation of CSLBs} 

Typically, CSLBs were made by spontaneous spreading and fusion of small
unilamellar vesicles (SUVs) on the particle surface. An SUV dispersion prepared
via either extrusion or sonication was mixed with the particles, allowing a
bilayer to spread on the surface for at least one hour. Subsequently the CSLBs
were washed to remove excess SUVs. We observed no substantial differences in
the obtained CSLBs between the two methods presented here.

\subsubsection*{CSLB preparation: method 1}

\SI{500}{\micro\gram} of a mixture of DOPE-Rhodamine (\SI{0.2}{mole \percent})
and varying amounts of POPC and PEGylated lipids was dried by two hours of
vacuum desiccation and then re-suspended to a \SI{2}{\gram\per\liter}
dispersion with HEPES buffer type 1. The solution was vortexed for
\SI{15}{\minute} to produce multilamellar vesicles. Then, the vesicle
dispersion was extruded 21 times with a mini extruder (Avanti Polar Lipids)
equipped with two \SI{250}{\micro\liter} gas-tight syringes (Hamilton), two
drain discs and one nucleopore track-etch membrane (Whatman). The pore size of
the membrane was 0.05 or 0.1 \si{\micro\meter} for experiments with
DOPE-PEG(2000) and DOPE-PEG(3000-5000), respectively. The as-prepared
\SI{50}{\micro\liter} of SUVs were added to \SI{1}{\milli\liter} of
\SI{0.05}{\percent w/v} of particles dispersed in HEPES buffer 1. The particles
were gently rotated for \SI{1}{\hour}. The resulting dispersion was centrifuged
at \SI{419}{rcf} for \SI{1}{\minute} and the supernatant replaced with HEPES
buffer type 1 to remove any SUVs present in the dispersion.

This method was used for all experiments regarding the influence of particle
material, surface roughness and the effect of polymer insertion on the
spreading and mobility of the lipid bilayer.

\subsubsection*{CSLB preparation: method 2}

Typically, a lipid mixture of \SI{98.9}{\mole\percent} DOPC,
\SI{1}{\mole\percent} DOPE-PEG(2000) and \SI{0.1}{\mole\percent} DOPE-Rhodamine
or TopFluor-Cholesterol in chloroform was dried overnight in a glass vial
covered with aluminum foil under vacuum desiccation. We investigated different
PEGylated lipid lengths and molar ratios. After drying, \SI{1}{\milli\liter}
HEPES buffer type 2 or 3 was added to reach a concentration of
\SI{2}{\gram\per\liter}. The dispersion was vortexed for \SI{30}{\minute},
after which it became turbid. It was then transferred to a plastic test tube
and ultrasonicated using a tip sonicator (Branson Sonifier SFX150) set to
\SI{30}{\percent} of its maximum amplitude for a total time of \SI{30}{\minute}
using a pulsed sequence (\SI{18}{\second} on/\SI{42}{\second} off, total on
time \SI{9}{\minute}) on ice to prevent heating. The SUV dispersion was then
centrifuged for \SI{45}{\minute} at \SI{2029}{rcf} to sediment larger vesicles
and titania particles originating from the tip \cite{Cremer1999}.
\SI{200}{\micro\liter} SUVs were taken from the top to isolate the smallest
vesicles.

\SI{17}{\micro\liter} of \SI{0.5}{\gram\per\liter} SUVs in HEPES buffer 2 or 3
were mixed with \SI{35}{\micro\liter} \SI{0.5}{\percent w/v} of particles in
Milli-Q water, leading to a surface ratio of SUVs:particles of 8:1. The
dispersion was gently rotated for \SI{1}{\hour}. The particles were centrifuged
at \SI{43}{rcf} for \SI{2}{\minute} and the supernatant was replaced with HEPES
buffer type 2 or 3 to remove any remaining free SUVs from the dispersion.
Alternatively, the particles were allowed to sediment by gravity for
\SI{30}{\minute} instead of centrifuging and the supernatant was replaced.

This method was used for all experiments regarding the colloidal stability of
CSLBs, the mobility of inserted DNA and the mobility of self-assembled CSLB
clusters.

\subsection*{Coating CSLBs with DNA for self-assembly}

After the particles were coated with a lipid bilayer using method 2, various
amounts of inert and/or linker DNA were added and the dispersion was gently
rotated for \SI{1}{\hour}. To remove any remaining free DNA strands in
solution, the particles were washed by centrifugation for \SI{2}{\minute} at
\SI{43}{rcf}, or alternatively, by sedimentation by gravity for
\SI{30}{\minute}, and the supernatant was replaced three times by HEPES buffer
type 2 or three.

We characterize the amount of dsDNA that we add as a surface density
$\sigma_{\mathrm{DNA}}$, which we define as \begin{align} \sigma_{\mathrm{DNA}}
&= \frac{N_{\mathrm{DNA}}}{A_{\mathrm{CSLB}}}, \label{eq:sigma_dna} \end{align}
    where $N_{\mathrm{DNA}}$ is the total number of dsDNA strands and
    $A_{\mathrm{CSLB}}$ is the total surface area of the CSLBs. The total
    number of dsDNA strands and particles were estimated from the reported
    stock concentrations. In this calculation, we assume that all the added
    dsDNA strands are distributed homogeneously over all particles and that no
    dsDNA remains in solution. We typically used
    $\sigma_{\mathrm{DNA}}=$~\SI{320}{\per\square\micro\meter} dsDNA linkers to
    obtain flexible structures.

Particle clusters were formed by mixing two particle types coated with
complementary DNA linkers in a 1:1 number ratio in a round metal sample holder
on a polyacrylamide coated cover slip (see \cite{Wel2016} for details). The
polyacrylamide coating keeps the particles from sticking to the glass surface,
allowing them to cluster via diffusion limited aggregation.

\subsection*{Sample characterization} 

The samples were imaged with an inverted confocal microscope (Nikon Ti-E)
equipped with a Nikon A1R confocal scanhead with galvano and resonant scanning
mirrors. A \SI{100}{\times} oil immersion objective ($\mathrm{NA} = 1.49$) was
used. A \SI{561}{\nm} laser was employed to excite the Lissamine Rhodamine dye,
a \SI{488}{\nm} laser was used to excite the TopFluor-Cholesterol dye. The
excitation light passed through a quarter wave plate to prevent polarization of
the dyes. 500-550~\si{\nm} and 565-625~\si{\nm} filters were used to separate
the emitted light from the TopFluor and the Rhodamine dyes, respectively.

The charge of the particles in MilliQ water was determined via zeta potential
measurements using a Malvern Zetasizer Nano ZS.

\subsection*{Fluorescence recovery after photobleaching (FRAP)} 

We used fluorescence recovery after
photobleaching (FRAP) to check the mobility of the lipids in a CSLB. A circular
area of the fluorescent sample was bleached, the recovery signal was collected and normalized as
\be
I_{\rm corr}(t) = \frac{I(t)}{I(t=0)I_{\rm ref}(t)}\;,
\label{eq:i_norm}
\ee
where $I_{\rm corr}(t)$ is the measured intensity $I(t)$ normalized with
respect to the intensity just before bleaching $I(t=0)$ and corrected for
bleaching through measurement of the intensity of a non-bleached reference
area, $I_{\rm ref}(t)$. Additionally, we subtracted the background signal from
$I$ and $I_{\rm ref}$. We found that the signal can be fitted using the following expression: 
\be
I_{\rm corr}(t) = A\left(1-e^{-\frac{t-t_0}{\tau}}\right)
\label{recovery}
\ee
where $A$ is the extent of the recovery, $t-t_{0}$ is the time
elapsed since the beginning of the recovery process and $\tau$ the recovery
time.

While there is a simple relation linking $\tau$ to $D$ for circular bleaching
areas on planar surfaces \cite{Axelrod1976}, we are not aware of a similar
expression for a spherical surface that is partly bleached from the side, as is
the case in our experiments. Therefore, we quantify the lateral mobility in
terms of the recovery time $\tau$ only.

All FRAP experiments on silica particles were performed using \SI{7}{\um}
particles, unless stated otherwise. To measure the lateral mobility of DNA
linkers using FRAP, no fluorescently labeled lipids were used and instead, we
used a high linker DNA concentration
($\sigma_{\mathrm{DNA}}=$~\SI{3e5}{\per\square\micro\meter}) that provided a
sufficiently bright fluorescent signal.

\subsection*{Particle stability analysis}
\label{sec:stab_tracking}

To estimate the colloidal stability of particles, we rotated the particles
(\SI{0.4}{\percent w/w}) in a test tube for at least \SI{1}{\hour}, thereby
allowing them to aggregate. We then immobilized some of the clusters on a glass
substrate, allowing us to take a ``snapshot'' of the cluster distribution at
that time. Particles were located in bright-field microscopy images of these
sedimented, semi-dilute (volume fraction $\phi\approx0.001$) samples. The
cluster sizes were determined by using the \texttt{bandpass}, \texttt{locate}
and \texttt{cluster} functions from TrackPy \cite{trackpy}. Erroneously tracked
positions were corrected manually. The separation distance below which
particles are considered to be part of a cluster was chosen to be $1.1D$, where
$D$ is the particle diameter. This can lead to a small overestimation of the
number of clusters when particles are close together but have not irreversibly
aggregated. We defined the fraction of single particles $f_{\mathrm{single}}$
as the number of detected clusters with a size of 1 (i.e. single particles)
divided by the total number of individual particles. The error on this
fraction was estimated as the standard deviation of the average cluster size
divided by the square root of the total number of particles. For each
measurement, we analyzed between \numrange{150}{4000} individual particles.

\subsection*{Trimer flexibility analysis}

We have analyzed three linearly-connected CSLBs that were functionalized with
inert dsDNA and linker dsDNA. To quantify the mobility of the self-assembled
trimers, we tracked the position of the three individual particles in
bright-field movies as a function of time and calculated the opening angle
$\theta$ between them. For tracking and calculating $\theta$, we used a custom
algorithm that is depicted in Figure S2 of the Supporting Information. 

First, the user selects the particles of interest from the first frame (see
Figure S2~A). This increases the computational efficiency of tracking because
it reduces the number of tracked features and allows for cropping of all
frames. We then iterate over all frames to identify the positions of the
selected particles. Each current frame is inverted, so that all particles have
a ring of high intensity around them (see Figure S2~B). The frame is converted
to polar coordinates with the current provisional particle position at the
origin (see Figure S2~C), where the provisional position is the one that the
user selected for the first frame and the previous tracked position for all
subsequent frames. For each row (each polar angle), the position of maximum
intensity is found (see Figure S2~D) and these coordinates are then converted
back to the original Cartesian coordinate system of the frame (see Figure
S2~E). A circle is fitted to these coordinates using a least squares method
(see Figure S2~F). After all three particles are found in this frame, the
opening angle between them is determined using simple trigonometry (see Figure
S2~G). From the opening angles of all the frames, we calculated the mean
squared displacement of the angle (MSAD or ``joint flexibility'' $J$)
\cite{Chakraborty2016a}.

We analyzed the free energy of trimers as function of opening angle using two
methods: 1) by converting the histogram to the free energy using Boltzmann
weighing and 2) using a maximum likelihood estimation method of angular
displacements \cite{Wel2016,Wel2017}. We confirmed that both methods agreed and
show only the result of the maximum likelihood method, because it allows us to
estimate the error in our measurement. We now describe these methods in detail.

\subsubsection*{Trimer free energy: Boltzmann weighing} 

We obtained a histogram of opening angles between \SIrange{60}{300}{\degree}
with a bin width of \SI{3}{\degree}. We then mirrored and averaged the data
around \SI{180}{\degree} and converted this to a probability density function.
From the probability density function we determined the free energy using
Boltzmann weighing, \begin{align} \frac{F}{k_B T} &= -\ln{P} + \frac{F_0}{k_B
T} \label{eq:boltzmann}, \end{align} where $F$ is the free energy, $k_B$ is the
Boltzmann constant, $T$ is the temperature, $P$ is the probability density and
$F_0$ is a constant offset to the free energy, which we chose at a reference
point (\SI{180}{\degree}) so that the free energy is equal to zero there.

\subsubsection*{Trimer free energy: maximum likelihood estimation} 

While the Boltzmann weighing method is very straightforward, it gives no
information about the experimental error. To estimate the error, we used an
analysis that is based on a maximum likelihood method in which particle
displacements are modelled \cite{Wel2016,Wel2017}, which we adapted for our
experimental system. We used (angular) displacements because for Brownian
particles they are uncorrelated in time, in contrast to positions (or values of
the opening angle). This means that using this method, we can obtain reliable
results even for a limited number of particles \cite{Sarfati2017}.

To summarize, we followed the method outlined in chapter 3.4.2 of
\cite{Wel2017}: we find the maximum likelihood estimate of the local force
field $F(\theta)$ by using a model for the transition probability $P$:
\begin{align}
    P(\theta_1,t+\tau|\theta_0,t)=&\left(4\pi D \tau\right)^{-\frac{1}{2}}\\\nonumber &\exp\left( -\frac{(\Delta\theta-\beta D F(\theta) \tau)^2}{4\pi D \tau} \right)
   \label{eq:transition_prob}
\end{align}
where $\theta_0$ is the opening angle at a time $t$ and $\theta_1$ is the angle
at a later time $t+\tau$, $\tau$ is the time between measurements, $D$ is the
diffusion coefficient determined from the mean squared displacement,
$\Delta\theta=\theta_1-\theta_0$ and $\beta$ is the Boltzmann constant times
the temperature. A Baysian method was used to find the maximum likelihood
estimate by using emcee \cite{emcee} and the error was determined as the
standard deviation of the chain of Markov Chain Monte Carlo (MCMC) samples. We
determined the free energy up to an arbitrary choice of a reference energy by
numerical integration of this force. This free energy was then mirrored around
\SI{180}{\degree} and averaged to determine the free energy between
\SI{60}{\degree} and \SI{180}{\degree}. We observed a boundary effect inherent
to the analysis for angles smaller than \SI{60}{\degree}$+\sqrt{2J\tau}$ (where
$J$ is the joint flexibility) leading to a slight overestimation of the free
energy for those angles.

\begin{figure}[ht]
    \centering
    \includegraphics[width=0.8\linewidth]{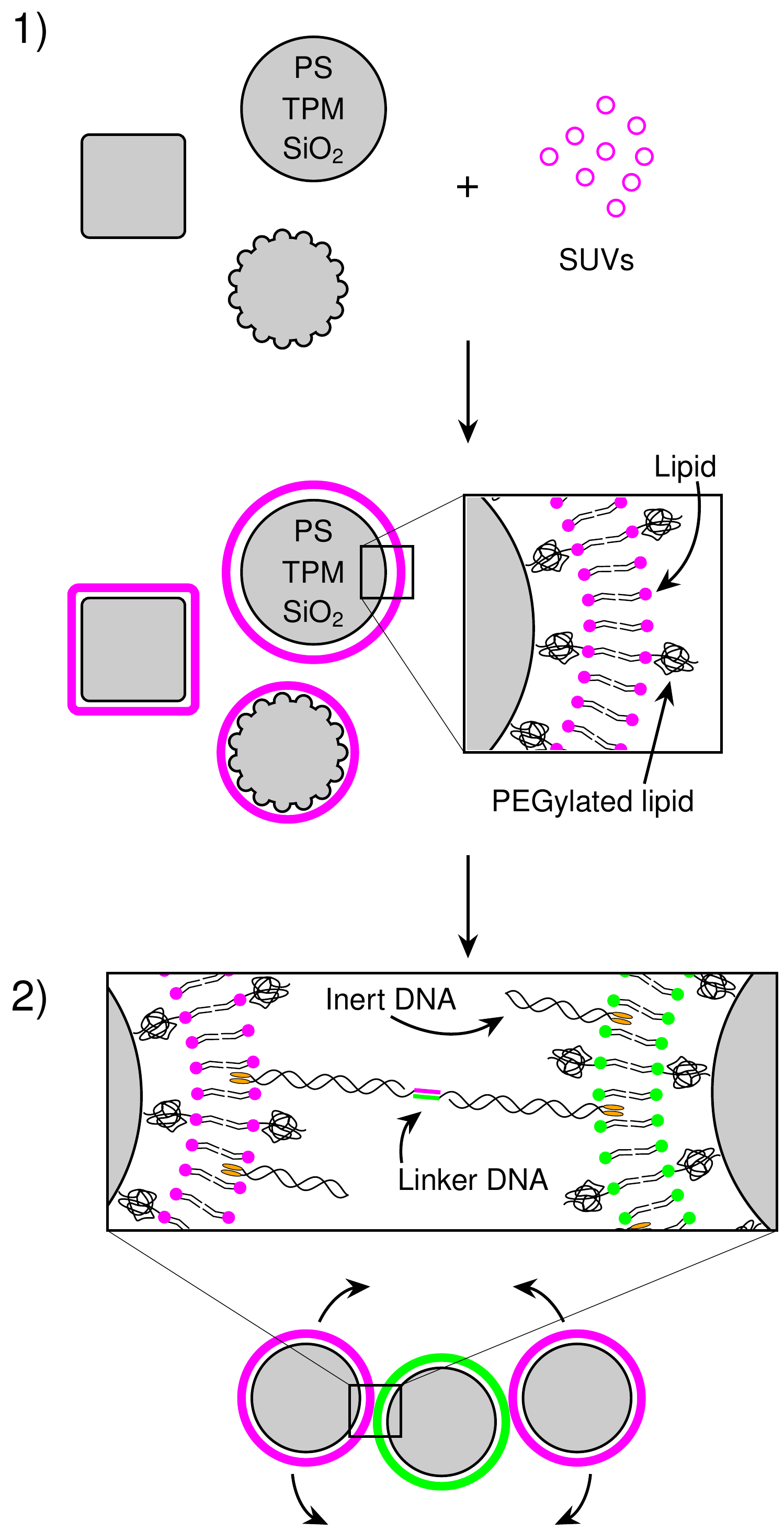}

    \caption{\textbf{Overview of the experimental system} \textbf{Step 1)}
        Micrometer sized colloidal particles are coated with a lipid bilayer by
        adding small unilamellar vesicles (SUVs) that rupture and spread on the
        particle surface. We varied the composition of the lipids, as well as
        the material and shape of the particles. \textbf{Step 2)} DNA linkers with
        hydrophobic anchors can be added to make particles that are
        functionalized with DNA with complementary sticky ends. When the lipid
        bilayer is fluid, the linkers can diffuse over the particle surface and
        therefore also the linked particles can slide over each other.
\label{fig:overview}} 
\end{figure}

\section*{Results and Discussion}

We will now characterize every step in the formation of CSLBs. In the first
section of the results we study the homogeneity and mobility of the lipid
bilayer on colloidal particles made from different materials, and therefore
various surface functionalities and degrees of roughness. Furthermore, we
investigate the effect of PEGylated lipids on the homogeneity and mobility of
the lipid bilayer and their use as steric stabilisers to prevent unspecific
aggregation. 

Having found conditions that yield colloidal particles with a homogeneous and
mobile bilayer, we subsequently introduce double-stranded DNA connected to a
hydrophobic anchor into the bilayer, as shown in \autoref{fig:overview}.2. We
employ DNA constructs both with and without single-stranded sticky ends, to
investigate their use in DNA-mediated binding and their effect on colloidal
stability, respectively. Finally, we demonstrate that CSLBs can be used for
self-assembly by employing DNA linkers with complementary single-stranded
sequences. We use FRAP to measure the lateral mobility of DNA linkers on the
particle surface inside and outside the bond area. In this way, we show that
they are mobile if the bilayer is fluid and that, in this case, the particles
can freely roll over each other's surfaces when bonded. 

\subsection*{Lipid bilayer coating of colloidal particles} 

To use CSLBs in self-assembly studies or as model membrane systems, it is
critical that a homogeneous and fluid bilayer forms on the colloidal particles.
This implies successful assembly of both leaflets of the bilayer and lateral
mobility of the lipids, and hence proteins, linkers, and larger lipid domains,
in the membrane. The formation of lipid bilayers on solid supports can be
achieved by deposition of SUVs under physiological conditions, as shown in
\autoref{fig:overview}.1. A combination of electrostatic and Van der Waals
forces lead to spreading and fusion of the liposomes on the surface of the
supports \cite{Richter2006, Sackmann1996, Raedler1995, GoZen}. Between the
surface of the support and the bilayer a thin layer of water remains, allowing
the lipids to laterally diffuse in the absence of other motion-restricting
forces. Previous studies on planar SLBs reported that there are many factors
which can prevent homogeneity and mobility of the bilayer \cite{Machan2010}.
These factors are related to the surface that is coated (its surface charge,
chemical composition and roughness), the aqueous medium (pH and ionic
strength), the SUVs (composition, charge, size, transition temperature) and the
temperature at which the lipid coating happens \cite{Richter2006, Jing2014}.
Here we will study how some of these factors, that are inherent to the use of
solid particles, influence the formation of supported lipid bilayers on
colloidal substrates.

\subsubsection*{Influence of the chemical properties of the particle surface} 

\begin{table*}[ht]\small
    \caption{Classification of bilayer spreading and mobility based on the material and the surface charge of the colloidal substrates. All Zeta potential measurements were performed in MilliQ water at room temperature. \label{table:colloidalsubstrates}}
\centering
\begin{tabularx}{0.95\linewidth}{X S[retain-explicit-plus] l l}
    \toprule
    \textbf{Material} & \textbf{Zeta potential [mV]} & \textbf{Homogeneous} & \textbf{Mobile}   \\ \midrule
    Silica spheres (St\"{o}ber method, Microparticles GmbH)  & -56\pm6 &  yes&  yes\\
    Hematite cubic particles \cite{Sugimoto1992}& +39\pm5 &  no &  no  \\
    Silica-coated hematite cubic particles \cite{Rossi2011} & -32\pm6 & yes&  yes    \\
    Polystyrene spheres (Sigma Aldrich) & -38\pm2 & no &  no   \\
    Polystyrene spheres with carboxyl groups \cite{Appel2013} & -43\pm1 & no &     no\\
    Silica-coated Polystyrene-TPM anisotropic particles \cite{Rinaldin2018} & -33\pm1   &  yes &  yes  \\
    TPM spheres  \cite{VanDerWel2017TPM} & -42\pm1 & yes   & no  \\
    TPM spheres with carboxyl groups \cite{DohertyTBP} & -46\pm1 &  no  & no  \\
    TPM spheres with amino groups \cite{DohertyTBP} & -12\pm4   &   no  & no \\
    \bottomrule
\end{tabularx}
\end{table*}

The available variety of colloids with anisotropic shapes makes them attractive
for self-assembly and model membrane studies. Current synthetic procedures
tailored to obtain different shapes, however, typically rely on the use of
specific materials and therefore yield colloids with different surface
properties. We have selected a range of particles of different shapes and
commonly used materials to test for membrane homogeneity and mobility after
coating with SUVs. In particular, we tested silica spheres prepared by a
sol-gel method, commercially available polystyrene spheres, polystyrene spheres
with carboxyl groups made using a surfactant-free dispersion polymerization
method \cite{Appel2013}, 3-(trimethoxysilyl)propyl methacrylate (TPM) spheres
\cite{VanDerWel2017TPM}, TPM spheres functionalized with carboxyl and amino
groups \cite{DohertyTBP}, silica-coated polystyrene-TPM spheres and symmetric
and asymmetric dumbbells \cite{Rinaldin2018}; as well as hematite cubes
\cite{Sugimoto1992} and silica-coated hematite particles \cite{Rossi2011}.
Silica-coated polystyrene-TPM dumbbells and hematite cubes were obtained by
depositing a silica layer following the St\"{o}ber method \cite{Castillo2014}. 

After coating, we visually inspect the lipid-coated particles using confocal
microscopy and consider bilayers to be homogenous if more than 50\% of the
particles do not show defects in the bilayer. We characterize the bilayer
fluidity by measuring the mobility of the fluorescently-labeled lipids on the
colloid surface using FRAP. After bleaching, we observe the recovery of the
fluorescence intensity due to the diffusion of the dyed lipids in and out of
the bleached area. We consider the lipids and thus the bilayer to be mobile if
the intensity signal recovers homogeneously in the bleached area, otherwise we
consider them to be (partially) pinned to the surface.

Our first observation was that only particles that possess a silica surface
exhibited homogeneous and mobile bilayers
(\autoref{table:colloidalsubstrates}). We did not succeed in coating colloids
made from polystyrene or hematite with a homogeneous bilayer. However, once
such substrates were first coated with a silica shell, the bilayer was found to
be both homogeneous and mobile. Unexpectedly, particles made from an
organosilica compound (TPM) whose surfaces are similar to silica
\cite{VanDerWel2017TPM} only showed homogeneous, but not mobile bilayers. Since
silica, TPM and polystyrene colloids were all negatively charged, we conclude
that the chemical composition of the substrate and not only the surface charge
plays a fundamental role in the homogeneity and fluidity of the bilayer. These
results agree with previous experiments on planar SLBs, in which silica-based
surfaces were found to be one of the most reliable supports for homogeneous and
mobile lipid bilayers \cite{Richter2006}.  

Since colloidal particles are often functionalized with different groups on the
surface, we furthermore have characterized the bilayer on particles equipped
with surface groups commonly used in colloidal science
(\autoref{table:colloidalsubstrates}). While TPM particles with an unmodified
silica-like surface showed homogeneous bilayers, we found that
functionalization with negatively charged carboxyl or positively charged amino
group prevented spreading and fusion of the lipid vesicles. Likewise,
functionalization of polystyrene spheres with carboxyl groups did not enhance
the homogeneity of the lipid bilayer. A previous study on planar SLBs reported
that the spreading of SUVs depends on the combination of the molecular ions in
the buffer and the type and density of surface charge \cite{Cha2006}. While
amino-functionalized surfaces are hence expected to disrupt the spreading of
SUVS in the presence of the negatively charged HEPES molecules of the buffer,
the observation of inhomogenous bilayers on carboxyl-functionalized surfaces
can likely be allocated to an insufficiently dense surface coverage. We
conclude that similar to planar SLBS, the homogeneity and fluidity of the
bilayer of CSLBs is dependant on a complex interplay of the chemical and
physical properties of the lipids and the particle's surface.

\subsubsection*{Influence of particle curvature differences} 

\begin{figure*}[ht]
    \centering
    \includegraphics[width=\linewidth]{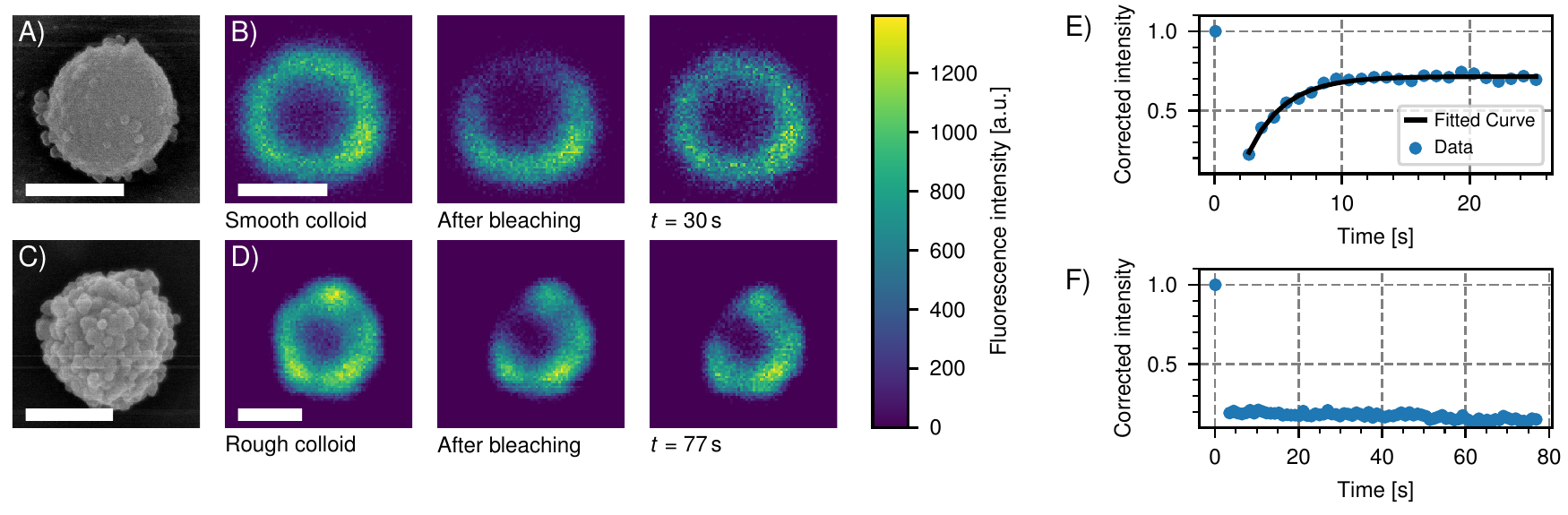}

\caption{\textbf{Effect of surface roughness on lipid bilayer formation.}
Scanning electron microscopy images (SEM) images of \textbf{A)} a smooth and
\textbf{C)} rough polystyrene-TPM particle coated with silica. Sequence of a FRAP
experiment before bleaching, just after bleaching and after 30s and 77s for the \textbf{B)}
smooth and \textbf{D)} rough particle, respectively. On the right, the fluorescence intensity as a function of time and
an exponential fit of the data for the \textbf{E)} smooth and \textbf{F)} rough 
particle are shown. The fluorescence recovery of the bilayer on the smooth particle shows
that the bilayer is fluid, in contrast to the rough particle which does not show a recovery of the fluorescence. Scale bars are \SIlist{1}{\micro\meter}.} 

\label{fig:roughness}
\end{figure*}

Another factor that may influence the successful formation of a homogeneous and
mobile bilayer is the variation in curvature of the colloidal substrate, which
may hinder spreading and fusion of SUVs. Curvature differences can originate
from the overall anisotropic shape of the particles or from surface roughness.
As discussed before, we found that particles with a comparably slowly varying
curvature, such as hematite cubes or symmetric and asymmetric dumbbells (see
\autoref{table:colloidalsubstrates}), had a fluid and homogeneous bilayer after
coating, if they featured a silica surface clean of any polymer residues from
synthesis. Particles with rough surfaces however have a much higher and
frequent variation in curvature. To investigate the effect of large curvature
differences, we prepared two batches of polystyrene particles which only
differed in their surface roughness and coated them with a silica layer
following a St\"{o}ber method \cite{Meester2016}. In \autoref{fig:roughness} we
show that particles with some roughness (A) can be homogeneously coated with a
bilayer (B) while particles with very rough surfaces (C) show an inhomogeneous
bilayer (D). FRAP experiments confirmed that the bilayer on the ``smooth''
surface is not only homogeneous, but also mobile, while the inhomogeneous
bilayer on the rough particle is immobile as indicated by the non-recovering
intensity signal. We conclude that the roughness of the surface plays an
important role in both bilayer homogeneity and mobility.

\subsubsection*{Influence of free and grafted polymers} 

\begin{figure}[ht]
    \centering
    \includegraphics[scale=1]{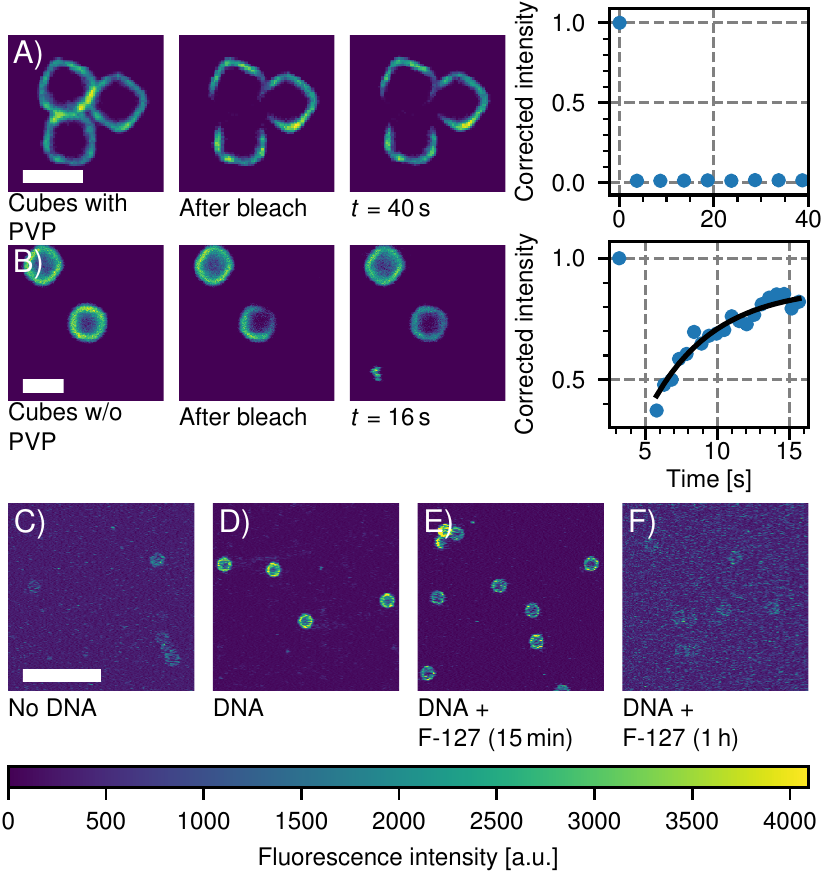}

\caption{\textbf{Effect of PVP on CSLB formation.} FRAP
experiment on a group of three cubes with \textbf{A)} and without \textbf{B)} PVP.
Only the sample without PVP shows recovery of the signal in the bleached area. Scale bars are \SIlist{2}{\micro\meter}.
\textbf{Influence of F127 in linker inclusion.} \textbf{C)} Control image of the fluorescence
        of the undyed CSLBs. \textbf{D)} Fluorescence intensity of CSLBs coated with
        \SI{20}{\nm} linker dsDNA. \textbf{E)} The same sample as in D), but
        imaged \SI{15}{\minute} after dispersion in
        \SI{5}{\percent} F-127 solution. The fluorescence on
        the particles was found to be significantly less homogeneous than in D).
        \textbf{F)} The same sample as in D), but imaged \SI{1}{\hour} after
    dispersion in \SI{5}{\percent} F-127 solution.
The fluorescence intensity is comparable to the uncoated control in C) so we
conclude that all dsDNA has been removed from the bilayer by F-127. Scale bar is \SIlist{10}{\micro\meter}.
\label{fig:PVP} } 
\end{figure}

Polymers or surfactants are often employed to stabilize colloidal particles in
solution \cite{DeGennes1987,Upadhyayula2012a,VanDerWel2017}, but may influence
the formation and mobility of the bilayer in CSLBs. Here, we test how the
presence of, for example, leftover polymers from particle synthesis, affects
the bilayer. We compare a sample of silica-coated hematite cubes with and
without PVP, a polymer commonly used in colloidal syntheses and conservation.
To remove the PVP from the surface after synthesis, we calcinated the colloids
at \SI{500}{\degreeCelsius} for 2h. \autoref{fig:PVP} shows that cubes with PVP
posses an inhomogeneous bilayer and the ones without it feature a bilayer that
homogeneously covers the surface (\autoref{fig:PVP} B). As expected for
St\"{o}ber silica surfaces, the bilayer on the colloids for which the PVP was
removed is also mobile, as indicated by the recovery of the fluorescence
intensity.

Moreover, the presence of polymers may not only affect the bilayer's
properties, but also the incorporation of functional groups such as DNA linkers
into it. We tested this by preparing CSLBs with fluorescently labeled DNA
linkers connected to double cholesterol anchors and transferring an aliquot of
this dispersion to a HEPES solution containing 5\% w/w of Pluronic F-127, a
polymer that is commonly used for the stabilization of colloidal particles.
While the fluorescent signal of the CSLBs with and without F-127 were initially
equal, already 15 minutes after mixing we observed less dsDNA fluorescence on
the CSLBs with F-127 compared to particles without it. After \SI{1}{\hour}, the
fluorescence intensity of the CSLBs with F-127 was comparable to that of
control particles not coated with linker dsDNA (\autoref{fig:PVP} C-F). We
therefore conclude that F-127 removed the cholesterol-anchored linker DNA from
the bilayer, in line with recent experiments on emulsion droplets coated with
mobile DNA linkers \cite{VanDerWel2018}.

\subsubsection*{Influence of PEGylated lipids on bilayer homogeneity and mobility} 

\begin{figure*}[ht]
    \centering
    \includegraphics[width=\linewidth]{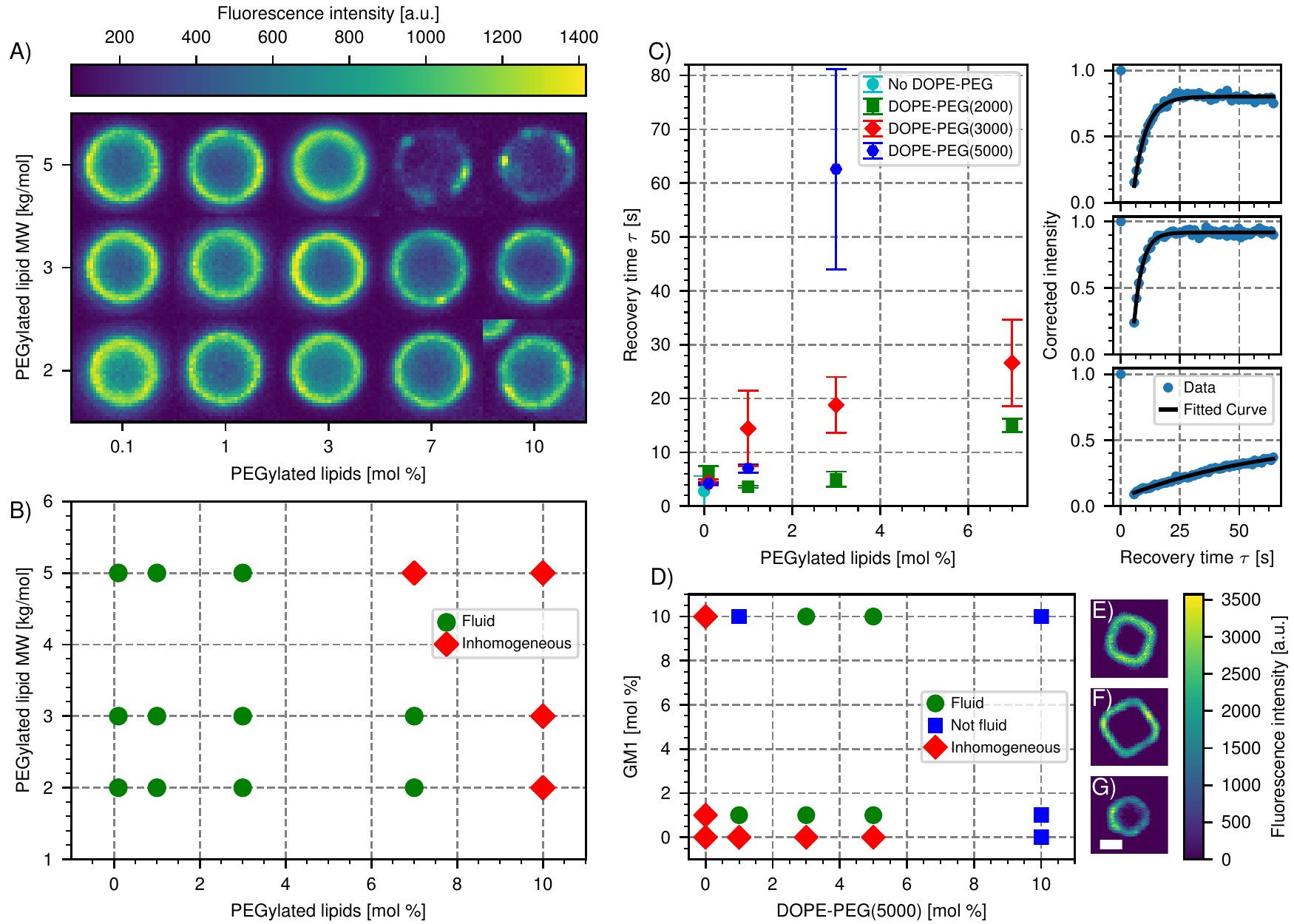}

\caption{\textbf{Effect of PEGylated lipids on CSLB formation and
fluidity.} \textbf{A)} Color-map of the intensity of the bilayer of spherical
CSLBs ordered by increasing molecular weight and concentration of PEGylated lipids. The images are
taken in the equatorial plane. \textbf{B)} Phase diagram of the state of
the bilayer for varying molecular weight and concentration of PEGylated lipids. \textbf{C)} Left: plot of the recovery time after FRAP depending on molar concentration and size of the PEGylated lipids. Right:
intensity recovery curves as a function of time from FRAP experiments of, from top to bottom, CSLBs without PEGylated lipids, with \SI{1}{\mole\percent} DOPE-PEG(2000) and \SI{3}{\mole\percent} DOPE-PEG(5000).
\textbf{D)} Phase diagram of spreading and fluidity of the bilayer on cubic silica shells depending on GM1 and PEGylated lipids. Colormap of a fluorescent image of a cubic bilayer made with \textbf{E)} \SI{1}{\mole\percent} GM1 and \SI{3}{\mole\percent} DOPE-PEG(5000), \textbf{F)} \SI{1}{\mole\percent} GM1 and \SI{10}{\mole\percent} DOPE-PEG(5000), \textbf{G)} no GM1 or DOPE-PEG(5000). Scale bar \SIlist{1.5}{\micro\meter}.
 \label{fig:PEG}} 
\end{figure*}

The presence of polymers in SLBs is not always detrimental, but may even
improve bilayer mobility. Previous studies on planar SLBs showed that membranes
can be supported by polymers covalently bound to lipids (lipopolymer tethers)
\cite{naumann2002polymer, Tanaka2005, Tanaka2006, Wagner2000, Deverall2008}.
Since lipopolymer tethers increase the thickness of the water layer between the
solid support and the bilayer \cite{naumann2002polymer, Tanaka2005, Tanaka2006}
they are thought to reduce the friction between the substrate and the bilayer,
allowing for higher diffusivity of lipid molecules and linkers.

Inspired by this, we study how a specific lipopolymer tether affects the
spreading and the fluidity of the bilayer in CSLBs. We used the lipopolymer
DOPE-PEG, a phospholipid with a covalently bound PEG molecule. We employed
PEGylated lipids with three different molecular weights: 2000, 3000 and 5000
\si{\gram\per\mole} in varying concentrations. It is important to note that
PEGylated lipids were introduced in the system during the SUV formation by
mixing them with the other lipids. This means that once the bilayer is formed,
they are present in both leaflets. 

We report in \autoref{fig:PEG}A-B the effect of varying concentration and
molecular weight of the lipopolymers on the spreading and the mobility of the
bilayer. In the absence of PEGylated lipids, the bilayer on the CSLBs was
observed to be fluid. At increased concentrations of DOPE-PEG, the bilayer
became inhomogeneous, which indicates insufficient spreading and fusion of the
SUVs.  This effect appeared at lower molar fractions for lipopolymers with
higher molecular weights of the DOPE-PEG. For completeness, we note that a
small fraction of particles in samples that are labeled as inhomogeneously
coated do exhibit a homogeneous, but nevertheless immobile, bilayer. We believe
that the reason for the observed inhomogeneity is two-fold. On the one hand,
higher concentrations of lipopolymers lead to an increased steric
stabilization, that prevents fusion of the SUVs and hinders the van der Waals
interactions between the SUVs and the substrate that aid spreading. On the
other hand, PEGylated lipids in the brush state increase the bending rigidity
of the SUV membrane, thereby preventing rupture and spreading on the surface
\cite{lipowsky1995bending}.

For fluid membranes, we quantified the mobility of the lipids by calculating
the recovery time from FRAP experiments, which is the time it takes a bleached
area to recover its fluorescence intensity. We find that the diffusion of the
lipids is faster for PEGylated lipids with a lower molecular weight and
increases with decreasing amount of the lipopolymers, see \autoref{fig:PEG}C.
This latter result agrees with a study performed on planar supported lipid
bilayers \cite{naumann2002polymer}. In the presence of lipopolymers, we find
the shortest recovery time (3.2 $\pm$ 0.02 s), e.g. highest diffusivity, for
\SI{1}{\mole\percent} of DOPE-PEG(2000). The decrease of the diffusion
coefficient with the amount of lipopolymer indicates that the PEGylated lipids
are pinned to the surface and in this way hinder the mobility of the other
lipids. 

We emphasize that the mobility of supported lipid bilayers in presence of
polymers is dependent on many factors and one may not extend our results to
other types of polymers, lipid bilayers or physiological environments
\cite{Deverall2008}. The complex interplay between polymers and the chemical
properties of the colloidal surface can lead to surprising results. For
example, and in contrast to what we reported above, we found that an homogenous
bilayer on cubic silica particles could only be obtained by using both
PEGylated lipids (DOPE-PEG(5000)) and a negatively charged lipid (GM1).
Interestingly, at high concentrations of PEGylated lipids the bilayer is very
homogeneous but not mobile (\autoref{fig:PEG}D). This is in contrast to silica
spheres coated with the same concentrations of lipopolymers and only
zwitterionic lipids, which do not possess a homogeneous bilayer, see
\autoref{fig:PEG}A. We indicated this state in which the bilayer is
homogeneous, but not fluid, with blue squares in \autoref{fig:PEG}D. A possible
origin of this unusual behavior could be the different porosity, surface
chemistry and charge of the silica cubes \cite{Castillo2014} compared to the
silica spheres (\autoref{table:colloidalsubstrates}).

\subsection*{Stabilizing CSLBs against aspecific aggregation}

To build specific colloidal structures from the bottom up, careful control over
the interactions between the particles is required. On the one hand, specific
attractive interactions may be employed to control which particles interact.
This specific binding can be achieved by using dsDNA linkers with complementary
sticky ends \cite{VanDerMeulen2013,Chakraborty2016a,Geerts2010}. On the other
hand, the particles need to be prevented from binding to each other
aspecifically: that is, not via dsDNA linker interactions but via other
attractive forces that act indiscriminately between all particles, such as Van
der Waals forces. In other words, it is crucial to be able to control the
colloidal stability of the CSLBs \cite{Valignat2005}.

In our experiments, stabilization by repulsive electrostatic interactions is
not a feasible route because surface charges are screened by the counterions in
the HEPES buffer that is needed to allow the complementary DNA sticky ends to
bind \cite{Geerts2010}. The ionic strength of the buffer must be higher than
\SI{50}{\milli\Molar} so that clusters are formed via DNA-mediated attractions
\cite{Biancaniello2007}. At these salt concentrations, even the bare silica
particles are no longer stabilized by their negatively charged surface groups.
Indeed, we found that both the bare silica particles and the silica particles
coated with a lipid bilayer aggregated in all buffers, as was previously
observed \cite{Nordlund2009}. The fraction of single particles determined from
light microscopy images was $f_{\mathrm{single}}=$~\num{0.67\pm0.10} for
uncoated silica particles after one hour of mixing in the buffer, while they
were previously stable in deionised water. We therefore explored different ways
to sterically stabilize the particles using higher concentrations of PEGylated
lipids, SDS and inert dsDNA strands.

\subsubsection*{Stabilization using SDS}

SDS is a surfactant with amphiphilic properties, consisting of a polar
headgroup and a hydrocarbon tail, that has been shown to stabilize emulsion
droplets coated with lipid monolayers \cite{Zhang2017}. Inspired by these
findings, we added SDS to the CSLBs after bilayer coating to increase their
stability. However, in contrast to lipid coated emulsion droplets we found no
significant increase in stability when we varied the SDS concentration between
\SIrange{0}{1}{\milli\Molar}. In fact, the highest concentration of
\SI{1}{\milli\Molar} led to a decrease in particle stability from
$f_{\mathrm{single}}=0.67$ without SDS to $f_{\mathrm{single}}=0.45$ at
\SI{1}{\milli\Molar}. This is likely caused by the disruptive effect that SDS
can have on lipid bilayers. In a study on large unilamellar vesicles (LUVs)
made from POPC, it was found that SDS can completely solubilize the vesicles
above concentrations of \SI{2}{\milli\Molar} \cite{Tan2002a}. While this
concentration is higher than the concentrations that we used here, we already
observed some damage to the bilayer. The resulting inhomogeneous coating may
allow aspecific ``sticking'' on patches that are not covered with a lipid
bilayer and a subsequent decrease in overall particle stability.

\subsubsection*{Stabilization using PEG}

\begin{figure}[ht]
\begin{center}
\includegraphics[width=0.95\linewidth]{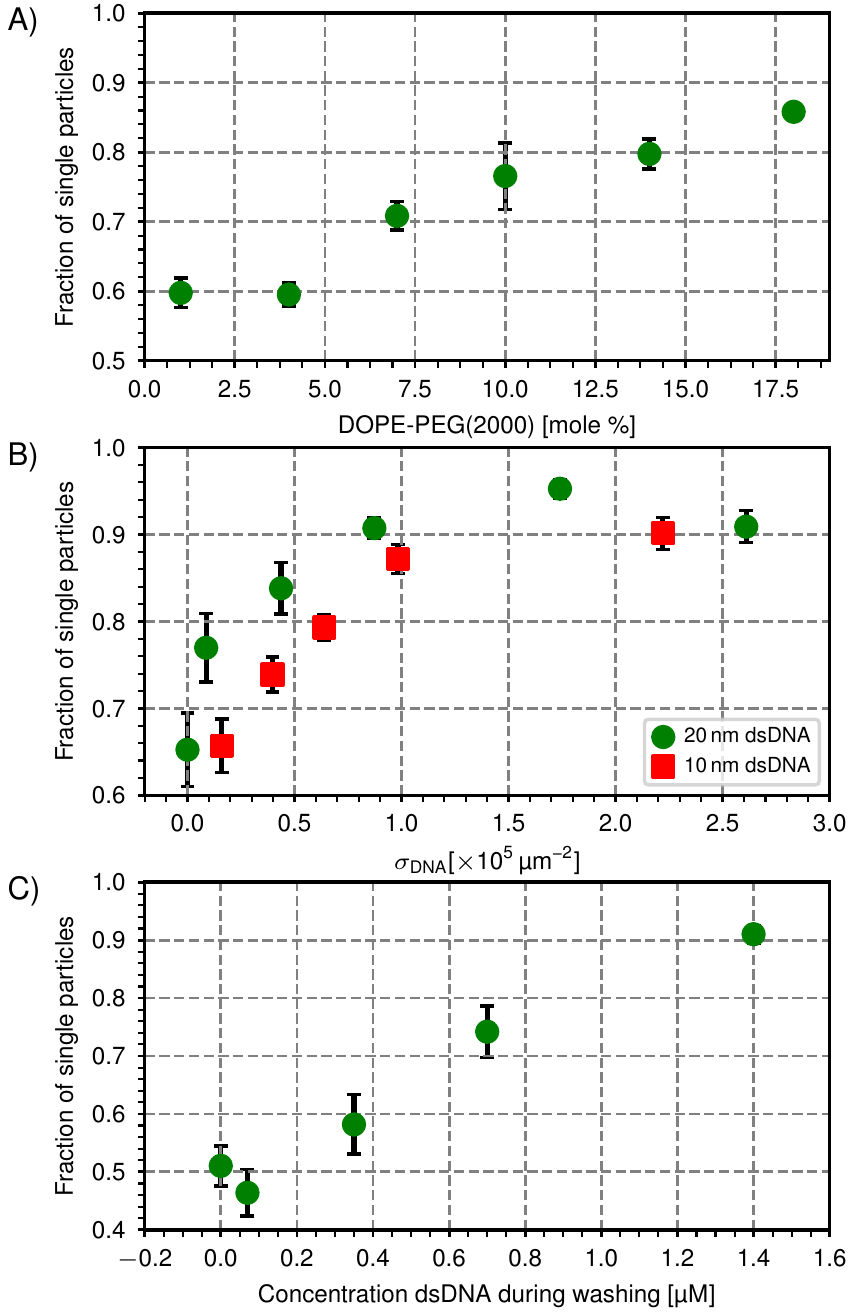}
\end{center}

\caption{\textbf{Steric stabilization of CSLBs.} \textbf{A)} Higher
    concentrations of DOPE-PEG(2000) lead to a higher fraction of single
    particles in the absence of linker DNA. Stability is shown after one
    washing cycle of \SI{2}{\minute} at \SI{43}{rcf} and overnight rotation. We
    hypothesize that above \SI{8}{\mole\percent}, the packing density of PEG on
    the surface of the membrane is high enough for the PEG to be in the brush
    state, making it more effective as a steric stabilizer. \textbf{B)}
    Increasing the number of dsDNA strands on the particle surface increases
    the particle's stability for two different lengths of inert dsDNA
    (\SI{10}{\nano\meter} and \SI{20}{\nano\meter}). \textbf{C)} Centrifugation and redispersion with a solution containing dsDNA affects the fraction of single particles. 
    After centrifuging particles
    that were initially stable ($f_{\mathrm{single}}=0.95$,
    $\sigma_{\mathrm{DNA}}=$~\SI{1.7e5}{\per\micro\meter\squared} dsDNA)
    $3\times$ at \SI{43}{rcf} for \SI{2}{\minute}, we observed aspecific
    aggregation ($f_{\mathrm{single}}=0.51$) in the absence of dsDNA strands in solution while increasing the concentration of dsDNA (\SI{20}{\nano\meter}) in the
washing solution could preserve stability.
\label{fig:stability}} 

\end{figure}

In contrast to SDS, PEGylated lipids can provide colloidal stability through
steric repulsions between the PEG moieties while also being easily integrated
into the bilayer through their lipid tails
\cite{DeGennes1987,Upadhyayula2012a,VanDerWel2017}. To test their use for
colloidal stabilization, we coated the particles with SUVs that contain a small
fraction of the following PEGylated lipids: DOPE-PEG(2000), DOPE-PEG(3000) and
DOPE-PEG(5000). Since we include these lipopolymers during SUV preparation,
they are part of both the inner and outer leaflet, as depicted in
\autoref{fig:overview}.1. At low concentrations, we observed no significant
change in the stability of the particles upon an increase in the concentration
of PEGylated lipids. For example, for DOPE-PEG(2000) stability remained
constant below \SI{5}{\mole\percent} as shown in \autoref{fig:stability}~A).
For concentrations between \SI{1}{\mole\percent} and \SI{7}{\mole\percent} for
DOPE-PEG(2000), \SI{1}{\mole\percent} and \SI{6.8}{\mole\percent} for
DOPE-PEG(3000) and \SI{2}{\mole\percent} and \SI{2.7}{\mole\percent} for
DOPE-PEG(5000), the average fraction of unclustered particles lay between
$f_{\mathrm{single}}=0.5$ and $f_{\mathrm{single}}=0.77$, with no clear trend
observed for different polymer lengths or concentrations. For all measurements,
we verified that the spreading of the SUVs was successful. We believe that the
stability of the particles did not improve significantly because at these
concentrations, the grafted PEG was in the mushroom state instead of the brush
state, and therefore not sufficient to provide steric stability
\cite{TanjaDrobek2005,Meng2004}. Therefore, we increased the concentration of
DOPE-PEG(2000) above \SI{7}{\mole\percent}. Indeed, as shown in
\autoref{fig:stability}, the colloidal stability increases above this
concentration, likely due to a transition from the mushroom to the brush state,
as was shown for similar lipopolymers (DSPE-PEG(2000)) around
\SI{8}{\mole\percent} \cite{Garbuzenko2005}.

While the colloidal stability can be improved by increasing the concentration
of PEGylated lipids, the bilayer is not fluid at the concentrations required
for colloidal stability as we showed earlier (see \autoref{fig:PEG}~B).
Therefore, embedded DNA linkers will also not be mobile in the bilayer and it
is not possible to form reconfigurable structures from CSLBs stabilized by
PEGylated lipids only.

\subsubsection*{Stabilization using inert dsDNA}
\label{sec:stability_dna}

Since PEGylated lipids cannot be used to provide steric stability because they
reduce the fluidity of the membrane, we explored an alternative route to
stabilize the CSLBs. Inspired by numerical findings that inert double stranded
DNA (dsDNA) can also act as a steric stabilizer via excluded volume
interactions between DNA strands on different particles
\cite{Angioletti-Uberti2014}, we employed dsDNA strands with a double
cholesterol anchor at one end to functionalize the CSLBs with this DNA via
hydrophobic insertion of the cholesterol into the bilayer
\cite{VanDerMeulen2014}, see \autoref{fig:overview}.2.

We varied the surface concentration $\sigma_{\mathrm{DNA}}$ (see
\autoref{eq:sigma_dna}) of two inert dsDNA strands with different lengths and
measured its effect on the particle stability, which is shown in
\autoref{fig:stability} B). The stability was determined after the particles
were coated with dsDNA and rotated for one hour. For both the
\SI{10}{\nano\meter} dsDNA and the \SI{20}{\nano\meter} dsDNA, we found that
increasing the number of grafted dsDNA strands per particle led to an increase
in particle stability from $f_{\mathrm{single}}=0.68$ without dsDNA to
$f_{\mathrm{single}}=$~\numrange{0.90}{0.95} above
$\sigma_{\mathrm{DNA}}=$~\SI{2e5}{\per\micro\meter\squared}. We found that the
\SI{20}{\nano\meter} dsDNA is slightly more efficient as a steric stabilizer
than the \SI{10}{\nano\meter} dsDNA, as can be seen in \autoref{fig:stability}
B). This is expected, because for the longer DNA, excluded volume interactions
between particles start to become important already at larger particle
separations than for the shorter DNA. Additionally, excluded volume
interactions between DNA strands on the same particle force the DNA to extend
outwards already at lower concentrations for the longer DNA strands as compared
to the shorter DNA strands. Therefore, the repulsion between the particles also
has a longer range, leading to better colloidal stability. However, at
concentrations above
$\sigma_{\mathrm{DNA}}=$~\SI{2e5}{\per\micro\meter\squared}, the difference
between the \SI{10}{\nm} and \SI{20}{\nm} dsDNA is less pronounced, because the
particles are so densely coated that adding longer or more DNA strands will not
stabilize the particles any further.

To use these particles in self-assembly studies, specific interactions need to
be present as well, which we here induce by adding dsDNA \textit{linkers}, that
is, dsDNA strands with a sticky end and double cholesterol anchors. After the
particles are functionalised, any excess linker DNA left in the solvent needs
to be removed to reduce unwanted aggregation or saturation of the
complementarily functionalized particles. To remove excess linker DNA, we
washed and redispersed the particles in buffer solution three times and
measured the particle stability afterwards. Unexpectedly, it decreased from
$f_{\mathrm{single}}=0.95$ before washing to $f_{\mathrm{single}}=0.51$ after
washing. To detect whether the partial removal of the stabilizing inert dsDNA
during washing had caused this aggregation, we washed the particles in a HEPES
buffer that contained various concentrations of inert dsDNA. As shown in
\autoref{fig:stability} C), increasing the concentration of dsDNA in the
washing solution led to an increase in particle stability after washing,
therefore confirming our hypothesis. Including \SI{1.4}{\micro\Molar} of inert
DNA in the washing solution allowed us to preserve the particle stability
($f_{\mathrm{single}}=0.91$). 

However, washing the particles with such high concentrations of dsDNA proved
detrimental to the bilayer and led to the formation of membrane tubes between
\SIrange{1}{3}{\um} long (see Figure S4 in the Supporting Information).
These membrane tubes are highly curved surfaces that are only formed under
specific conditions, for example a difference in spontaneous curvature between
the inner and the outer membrane leaflets \cite{Lipowsky2013}. In our case,
since dsDNA is only added after the formation of the bilayer it is only present
in the outer leaflet and hence induces a difference in spontaneous curvature
which causes formation of tubes. If the DNA would be present on both leaflets,
no tube formation is expected. We tested this by mixing the dsDNA with the SUVs
before coating the particles but found no fluorescence signal on the particles'
surface, implying the absence of a bilayer coating. We believe that the dsDNA
sterically stabilizes the SUVs and prevents spreading and fusion of SUVs on the
particle surface. 

To summarize, we found that inert double-stranded DNA can impart colloidal
stability to the CSLBs. Asymmetric distribution of DNA causes the formation of
membrane tubes at high coating concentrations (above approximately
\SI{1e5}{\per\square\micro\meter}), but is unavoidable if colloidal stability
needs to be preserved during repeated washing cycles.

\subsection*{Linker functionalization for self-assembly}

\begin{figure*}[ht]

\begin{center}
    \includegraphics[scale=0.95]{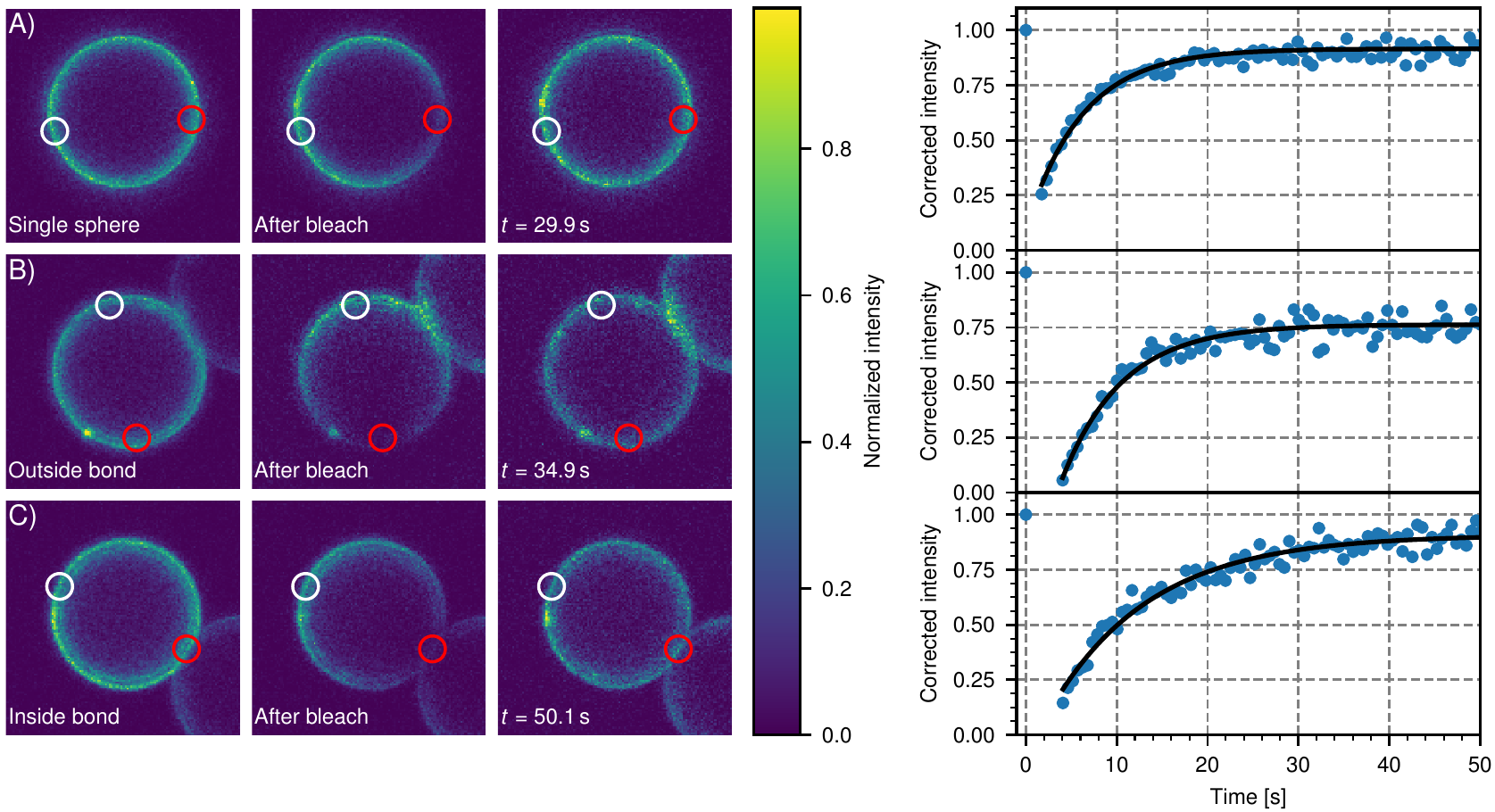}
\end{center}

\caption{\textbf{DNA linker mobility on CSLBs.} Three representative FRAP
    experiments are shown that probe the mobility of DNA linkers on the surface
    of a \SI{7}{\um} CSLB. In the intensity color plots, a red circle
    indicates the bleached area and a white circle shows the reference
    area. For all experiments, the DNA linkers are mobile.
    \textbf{A)} FRAP on a single sphere. We measured an average recovery time
    of \SI{8\pm2}{\second} from independent measurements on 16 different
    particles. \textbf{B)} FRAP on a sphere in a cluster, taken outside the bond
    area. We measured an average recovery time of \SI{6\pm2}{\second} from
    independent measurements on 15 different clusters. Note that, within the error, this is the same as the recovery time we measured for single
    particles in A). \textbf{C)} FRAP on a sphere in a cluster, inside the bond
    area. We measured recovery times of \SIrange{5}{62}{\second} with an
    average recovery time of \SI{17\pm16}{\second} from independent
    measurements on 12 different clusters. Note that this is longer than the
    recovery times we measured outside the bond area in A) and B), which
    indicates the diffusion is slower in the bond area. However, the spread in
    recovery times was also larger for bleached areas inside the bond area.
\label{fig:DNA_mobility}} 

\end{figure*}

To be able to employ CSLBs in self-assembly experiments, we induce specific
attractive interactions by using two sets of dsDNA linkers with complementary
single-stranded sticky ends that can form bonds via hybridization. We use two
hydrophobic anchors (either cholesterol or stearyl) per dsDNA complex to insert
the dsDNA linker into the outer leaflet (see \autoref{fig:overview}.2), while
at the same time adding inert dsDNA strands for steric stabilization. We use
double hydrophobic anchors because dsDNA with a single hydrophobic anchor is
less strongly confined to the bilayer \cite{VanDerMeulen2014}. The dsDNA is
attached to the double hydrophobic anchor with a tetraethylene glycol (TEG) or
hexaethylene glycol (HEG) spacer to allow it to swivel with respect to the
anchor. We label the dsDNA linkers with a fluorescent dye to image them using
confocal microscopy. Full details on the DNA strands we used can be found in
Table S1 of the Supporting Information.

Previous experiments \cite{Chakraborty2016a} have shown that several
interesting structures, such as flexible colloidal polymers and molecules, can
be obtained via self-assembly of CSLBs. In order to form these reconfigurable
structures, not only the lipids in the bilayer should be mobile but also the
grafted linker DNA should be mobile on the surface of the membrane. We can
quantify the mobility of dsDNA on the surface of the bilayer by measuring the
FRAP of fluorescently labeled DNA. Note that for these experiments we did not
employ fluorescent lipids. For a successful recovery after bleaching of the DNA
linkers in the binding patch between two particles, two requirements need to be
fulfilled: the DNA linkers outside of the binding patch have to be mobile and
the bleached linkers inside the binding patch have to be able to unbind, to
allow unbleached linkers to diffuse into the binding patch. We calculate the
melting temperature of the sticky end using an approximate formula
\cite{Nakano1999} and find a melting temperature $T_m =
$~\SI{30.4}{\degreeCelsius}, meaning that at \SI{25}{\degreeCelsius} the
probability for the sticky end to form a duplex is
$P($\SI{25}{\degreeCelsius}$)=0.68$ based on melting temperature considerations
only. Therefore, the sticky ends continuously bind and unbind in our
experiments, making fluorescence recovery possible, while the sheer number of
linkers in the patch area keeps the particles always bound.

We confirmed the mobility of linker DNA on the particle surface using FRAP
experiments, shown in \autoref{fig:DNA_mobility}. Note that the whole cluster
is immobilized on the (non-passivated) glass coverslip to enable FRAP on these
micrometer-sized particles. Therefore these clusters do not show translational
diffusion or cluster flexibility. In \autoref{fig:DNA_mobility} A), a
representative FRAP experiment on a single sphere is shown. We measured an
average recovery time of \SI{8\pm2}{\second} from independent measurements on
16 different particles. Outside the bond area (\autoref{fig:DNA_mobility} B),
we measured an average recovery time of \SI{6\pm2}{\second} from independent
measurements on 15 different clusters. Within the error, the diffusion of DNA
outside of the bond area is the same as on free particles as expected. Inside
the bond area (\autoref{fig:DNA_mobility} C), we measured recovery times of
\SIrange{5}{62}{\second} with an average recovery time of \SI{17\pm16}{\second}
from independent measurements on 12 different clusters. This is longer than the
recovery times we measured outside the bond area and on single particles, which
indicates the diffusion is slower in the bond area. However, the recovery time
inside the bond area varied greatly between different clusters indicating that
diffusion into and out of the bond area can be different. A likely cause is the
spread in the DNA concentration between individual particles in a batch. Higher
DNA concentrations imply a higher concentration of bonded linkers in the bond
area. This will sterically hinder the diffusion of unbonded linkers inside the
patch area and will thus lead to longer recovery times. Furthermore, we
hypothesize that the linker concentration in the patch area slowly increases as
a function of time after binding, so that the recovery time depends on the time
that has elapsed after the formation of the cluster, which we did not control.
In summary, we have shown here that the dsDNA linkers are mobile on each part
of the (un)bound particle, which is a prerequisite for creating flexibly linked
clusters.

\subsection*{Mobility of self-assembled structures}

\begin{figure}
\includegraphics[scale=0.95]{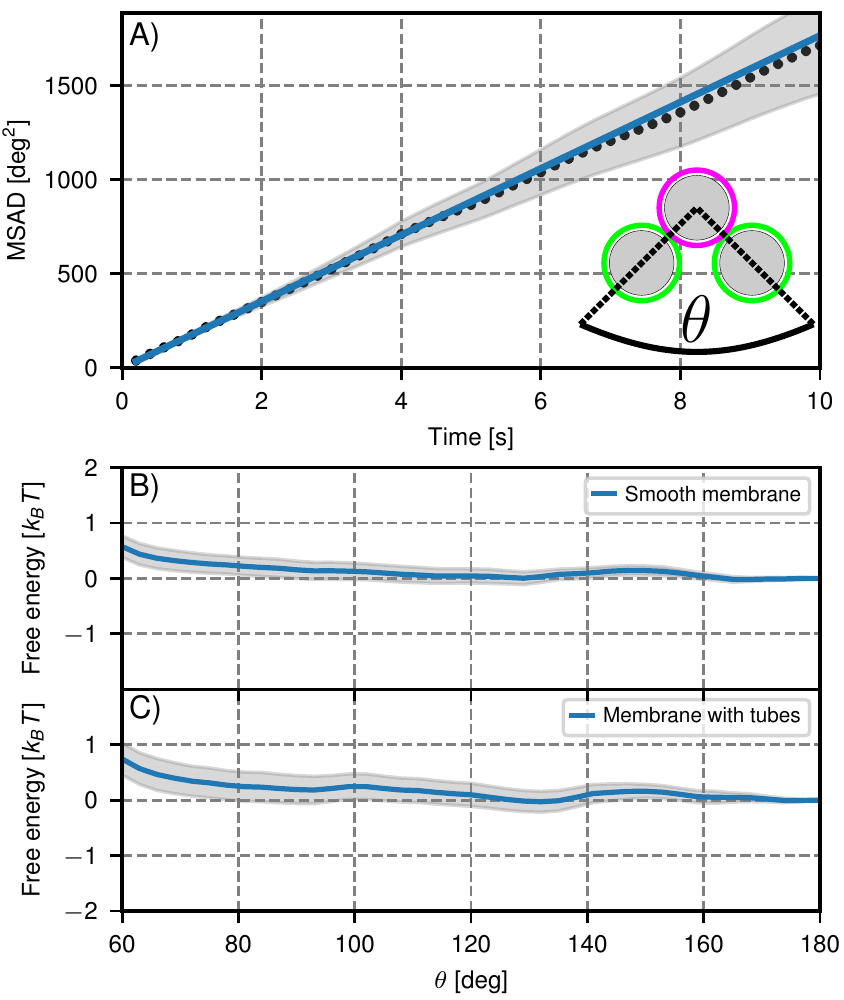}

\caption{\textbf{Mobility of self-assembled trimers} \textbf{A)} Mean squared
    angular displacement (MSAD) of the opening angle $\theta$ for a mobile
    trimer. The MSAD is linear and we find a joint flexibility
    $J=$~\SI{176\pm12}{deg^2\per\second} ($J=$~\SI{184\pm101}{deg^2\per\second}
    when averaging over all experiments). \textbf{B, C)} Free energy as function of
    opening angle $\theta$ for particles with \textbf{B)} smooth bilayers and
    \textbf{C)} bilayers that have membrane tubes. The grey shaded area marks one standard deviation
    confidence interval. We analyzed 53
    clusters with smooth bilayers and 18 clusters with membrane tubes. In both
    cases, we find no preference for specific opening angles within the experimental error, meaning the particles are
freely-jointed. Note that the slight repulsion at small angles is caused by
boundary effects inherent to our analysis method.\label{fig:trimer_theta}}

\end{figure}

The mobility of individual dsDNA linkers on the surface does not necessarily
imply that bonded clusters of DNA functionalized CSLBs are also reconfigurable.
For example, for emulsion droplets functionalized with DNA linkers, the large
linker patch that is formed between particles can slow down the motion when the
supporting fluid is inhomogeneous \cite{VanDerWel2018} and colloidal joints
lose their flexibility with an increasing concentration of dsDNA linkers in the
bond area \cite{Chakraborty2016a}.

To measure the flexibility of larger structures, we assembled CSLBs with
complementary dsDNA linkers and imaged chains of three CSLBs, so called
trimers, over time. We extracted the position of the individual particles and
the opening angle $\theta$ (see inset \autoref{fig:trimer_theta}~A) for all
frames and calculated the mean squared angular displacement (MSAD) to
characterize the flexibility \cite{Chakraborty2016a}. To investigate the
influence of the membrane homogeneity on the structural flexibility, we
compared trimers assembled from CSLBs with homogeneous, fluid bilayer to
trimers with bilayers that had spontaneously formed membrane tubes. In the
following, we will only show the free energy landscape for $\theta$ from
\SIrange{60}{180}{\degree} due to the symmetry of a trimer.

For trimers made from CSLBs with smooth lipid bilayers, we found that the
particles ($D=$~\SI{2}{\micro\meter}) move with respect to each other over the
full angular range. We analyzed the opening angle $\theta$ for \num{53}
different clusters by tracking the individual particles and calculating
$\theta$ for all frames (see inset \autoref{fig:trimer_theta}~A). The average
value of the flexibility of the trimers is
$J=$~\SI{184\pm101}{deg^2\per\second} (or
\SI{0.03\pm0.02}{\micro\meter\squared\per\second}) and agrees well with
previous experiments \cite{Chakraborty2016a}. The spread in the flexibility
that we observe is likely caused by the spread in dsDNA linker density on the
particles. We then determined the free energy using the maximum likelihood
estimation of angular displacements method (see Experimental Section), as shown
in \autoref{fig:trimer_theta}~B). We found no preference in the opening angle
that is significant with respect to the thermal energy $k_B T$, indicating that
the surface is smooth enough to allow the particles to move over one another
without restrictions. We observed a boundary effect inherent to the analysis
for angles smaller than \SI{60}{\degree}$+\sqrt{2J\tau}\approx$~\SI{69}{\degree}
(where $J$ is the joint flexibility) leading to a slight overestimation of the
free energy for those angles. In previous experiments, we observed a small
preference (0.9 $K_B T$ free energy difference) in opening angle around
\SI{140}{\degree} \cite{Chakraborty2016a}. We do not find the same preference
in angle for the experiments presented here. When re-analyzing the data from
\cite{Chakraborty2016a} using our maximum likelihood method, we noted that the
observed free energy difference is within the experimental error (see Figure S3
in the Supporting Information) and therefore we conclude that there is no
preference in opening angle. This is also in line with recent experiments on
flexibly linked emulsion droplets, which are freely-jointed like the CSLBs we
present here \cite{McMullen2018}.

To test the robustness of this structural flexibility in the presence of
membrane inhomogeneities, we prepared particles with a large number of membrane
tubes by washing with inert dsDNA. As shown in \autoref{fig:trimer_theta}~C),
this did not alter the flexibility of the clusters: they still expore the full
angular range and do not show a significant preferred opening angle, similar to
the particles coated with a smooth membrane. Surprisingly, this means that
particles stabilized by dsDNA can in principle be used for self-assembly
studies despite the fact that the high concentration of dsDNA causes membrane
tubes to form, because the tubes do not significantly alter the relative motion
of clustered particles.

\section*{Summary and Conclusions} 

We investigated various factors in the preparation protocol of colloidal
supported lipid bilayers (CSLBs) in view of their emerging use in self-assembly
and model membrane studies. Specifically, we focused on realizing a homogeneous
and fluid bilayer, while achieving colloidal stability and functionalization
with DNA linkers at the same time. 

Similar to what has been reported for flat supported lipid bilayers, we found
that the quality of the lipid bilayer on colloidal particles critically depends
on the material of the particle's surface. The bilayer was not fluid on
particles made from polystyrene (with or without carboxyl groups), hematite and
TPM particles (with or without carboxyl or amino groups). Colloids featuring a
silica surface, on the other hand, were able to host a fluid and homogenous
bilayer, at least in the absence of any polymer residues from the synthesis. We
furthermore observed that the variation in the substrate curvature does not
affect the bilayer formation if it is sufficiently gentle, while excessive
surface roughness can hinder the spreading and fusion of SUVs. 

Use of PEGylated lipids in the bilayer increased the colloidal stability, but
affected the bilayer homogeneity and mobility negatively. Addition of the
amphiphilic surfactant SDS led to a disintegration of the bilayer. A better way
to provide colloidal stability is by steric stabilisation by excluded volume
effects through the insertion of double-stranded inert DNA. Increasing the
concentration of dsDNA leads to an increase in colloidal stability.

Finally, we demonstrated that these CSLBs can be functionalized with
surface-mobile DNA linkers and assembled them into flexible structures of
freely-jointed particles. We found that local bilayer inhomogeneities in the
form of membrane tubes do not affect the free energy landscape of the connected
particles. 

CSLBs with fluid, homogeneous membranes and surface-mobile binding groups
have great promise in a wide range of applications and fundamental studies.
The fact that the bonded particles can flexibly move with respect to each other
opens the door to overcoming equilibration issues previously encountered in
hit-and-stick processes and assembling structures with internal deformation
modes. This enables studying the impact of structural flexibility on the phase
behaviour, such as the formation of crystals with new lattices or properties
\cite{Kohlstedt2013, Ortiz2014, Smallenburg, Hu2018}, and the experimental
realization of information elements for wet computing \cite{Phillips2014a}.
CSLBs with increased membrane fluidity also mimic biological membranes more
closely which may be advantageous for model membrane and cell biology studies
\cite{Sackmann2007a, Madwar2015, Mashaghi2013, Rinaldin2018, Fonda2018} smart drug
delivery \cite{Carmona-Ribeiro2012, Li2014, Savarala2010} and bio-sensing
applications \cite{Castellana2006, Chemburu2010}.

\paragraph*{Supporting Information} The Supporting Information consists of 4
figures and 1 table: large fields of view of microscopy images; a table listing
all employed DNA strands; a schematic representation of the particle tracking
algorithm; a graph showing data from a previous paper re-analyzed using a
different method; and a microscopy picture of three CSLBs that feature membrane
tubes.

\paragraph*{Acknowledgments} We thank Rachel Doherty and Vera Meester for
help with the colloidal syntheses and electron microscopy imaging, and Piermarco Fonda and Luca Giomi for useful discussions. This
work was supported by the Netherlands Organization for Scientific Research
(NWO/OCW), as part of the Frontiers of Nanoscience program and VENI grant
680-47-431. This project has received funding from the European Research
Council (ERC) under the European Union's Horizon 2020 research and
innovation program (grant agreement no. 758383).

\paragraph*{Author Contributions} MR and RWV contributed equally to the
work. MR and RWV performed the experiments and the data analysis. IC
performed exploratory experiments. MR, RWV and DJK conceived the
experiments and wrote the paper.

{\small\bibliography{references}{}}

\begin{thebibliography}{10}

\bibitem{Troutier2007}
Troutier, A.~L.; Ladavi{\`{e}}re, C.
\newblock {An overview of lipid membrane supported by colloidal particles}.
\newblock \emph{Adv. Colloid Interface Sci.} \textbf{2007}, 133~(1), 1--21.

\bibitem{Carmona-Ribeiro2012}
Carmona-Ribeiro, A.~M.
\newblock Preparation and Characterization of Biomimetic Nanoparticles for Drug
  Delivery.
\newblock In Nanoparticles in Biology and Medicine: Methods and Protocols,
  283--294. Humana Press, Totowa, NJ, \textbf{2012}.

\bibitem{Li2014}
Li, J.; Wang, X.; Zhang, T.; Wang, C.; Huang, Z.; Luo, X.; Deng, Y.
\newblock A review on phospholipids and their main applications in drug
  delivery systems.
\newblock \emph{Asian Journal of Pharmaceutical Sciences} \textbf{2015},
  10~(2), 81--98.

\bibitem{Savarala2010}
Savarala, S.; Ahmed, S.; Ilies, M.~A.; Wunder, S.~L.
\newblock {Formation and Colloidal Stability of DMPC Supported Lipid Bilayers
  on \ce{SiO2} Nanobeads}.
\newblock \emph{Langmuir} \textbf{2010}, 26~(14), 12081--12088.

\bibitem{Castellana2006}
Castellana, E.~T.; Cremer, P.~S.
\newblock {Solid supported lipid bilayers: From biophysical studies to sensor
  design}.
\newblock \emph{Surf. Sci. Rep.} \textbf{2006}, 61~(10), 429--444.

\bibitem{Chemburu2010}
Chemburu, S.; Fenton, K.; Lopez, G.~P.; Zeineldin, R.
\newblock {Biomimetic Silica Microspheres in Biosensing}.
\newblock \emph{Molecules} \textbf{2010}, 15~(12), 1932--1957.

\bibitem{Brouwer2015}
Brouwer, I.; Giniatullina, A.; Laurens, N.; Weering, J. R. T.~V.; Bald, D.;
  Wuite, G. J.~L.; Groffen, A.~J.
\newblock Direct quantitative detection of Doc2b-induced hemifusion in
  optically trapped membranes.
\newblock \emph{Nat. Commun.} \textbf{2015}, 6, 8387.

\bibitem{Sackmann2007a}
Sackmann, E.
\newblock {Supported membranes: Scientific and practical applications}.
\newblock \emph{Science} \textbf{2007}, 271~(5245), 43--48.

\bibitem{Madwar2015}
Madwar, C.; Gopalakrishnan, G.; Lennox, R.~B.
\newblock {Interfacing living cells and spherically supported bilayer lipid
  membranes}.
\newblock \emph{Langmuir} \textbf{2015}, 31~(16), 4704--4712.

\bibitem{Mashaghi2013}
Mashaghi, S.; Jadidi, T.; Koenderink, G.; Mashaghi, A.
\newblock {Lipid Nanotechnology}.
\newblock \emph{Int. J. Mol. Sci.} \textbf{2013}, 14, 4242--4282.

\bibitem{Rinaldin2018}
{Rinaldin}, M.; {Fonda}, P.; {Giomi}, L.; {Kraft}, D.~J.
\newblock {Geometric pinning and antimixing in scaffolded lipid vesicles}.
\newblock \emph{ArXiv e-prints} \textbf{2018}.

\bibitem{Fonda2018}
{Fonda}, P.; {Rinaldin}, M.; {Kraft}, D.~J.; {Giomi}, L.
\newblock {Interface geometry of binary mixtures on curved substrates}.
\newblock \emph{ArXiv e-prints} \textbf{2018}.

\bibitem{VanDerMeulen2013}
{Van Der Meulen}, S.~A.; Leunissen, M.~E.
\newblock {Solid colloids with surface-mobile DNA linkers}.
\newblock \emph{J. Am. Chem. Soc.} \textbf{2013}, 135~(40), 15129--15134.

\bibitem{VanDerMeulen2014}
{Van Der Meulen}, S. A.~J.; Dubacheva, G.~V.; Dogterom, M.; Richter, R.~P.;
  Leunissen, M.~E.
\newblock {Quartz crystal microbalance with dissipation monitoring and
  spectroscopic ellipsometry measurements of the phospholipid bilayer anchoring
  stability and kinetics of hydrophobically modified DNA oligonucleotides}.
\newblock \emph{Langmuir} \textbf{2014}, 30~(22), 6525--6533.

\bibitem{VanderMeulen2015}
Van~der Meulen, S. A.~J.; Helms, G.; Dogterom, M.
\newblock {Solid colloids with surface-mobile linkers}.
\newblock \emph{J. Phys.: Condens. Matter} \textbf{2015}, 27~(23), 233101.

\bibitem{Chakraborty2016a}
Chakraborty, I.; Meester, V.; Van~der Wel, C.; Kraft, D.~J.
\newblock {Colloidal joints with designed motion range and tunable joint
  flexibility}.
\newblock \emph{Nanoscale} \textbf{2017}, 9, 7814--7821.

\bibitem{Feng2013a}
Feng, L.; Pontani, L.-L.; Dreyfus, R.; Chaikin, P.; Brujic, J.
\newblock {Specificity, flexibility and valence of DNA bonds guide emulsion
  architecture}.
\newblock \emph{Soft Matter} \textbf{2013}, 9~(41), 9816--9823.

\bibitem{McMullen2018}
McMullen, A.; Holmes-Cerfon, M.; Sciortino, F.; Grosberg, A.~Y.; Brujic, J.
\newblock {Colloidomers: freely-jointed polymers made of droplets}.
\newblock \emph{ArXiv e-prints} \textbf{2018}.

\bibitem{Kohlstedt2013}
Kohlstedt, K.~L.; Glotzer, S.~C.
\newblock {Self-assembly and tunable mechanics of reconfigurable colloidal
  crystals}.
\newblock \emph{Phys. Rev. E - Statistical, Nonlinear, and Soft Matter Physics}
  \textbf{2013}, 87~(3).

\bibitem{Ortiz2014}
Ortiz, D.; Kohlstedt, K.~L.; Nguyen, T.~D.; Glotzer, S.~C.
\newblock {Self-assembly of reconfigurable colloidal molecules}.
\newblock \emph{Soft Matter} \textbf{2014}, 10~(20), 3541--3552.

\bibitem{Smallenburg}
Smallenburg, F.; Filion, L.; Sciortino, F.
\newblock {Erasing no-man's land by thermodynamically stabilizing the
  liquid-liquid transition in tetrahedral particles}.
\newblock \emph{Nat. Phys.} \textbf{2014}, 10~(9), 653--657.

\bibitem{Hu2018}
Hu, H.; Ruiz, P.~S.; Ni, R.
\newblock {Entropy Stabilizes Floppy Crystals of Mobile DNA-Coated Colloids}.
\newblock \emph{Phys. Rev. Lett.} \textbf{2018}, 120~(4), 048003.

\bibitem{Joannopoulos1997}
Joannopoulos, J.~D.; Villeneuve, P.~R.; Fan, S.
\newblock {Photonic crystals: putting a new twist on light}.
\newblock \emph{Nature} \textbf{1997}, 386~(6621), 143--149.

\bibitem{Lin2005}
Lin, Y.; Herman, P.~R.; Valdivia, C.~E.; Li, J.; Kitaev, V.; Ozin, G.~A.
\newblock {Photonic band structure of colloidal crystal self-assembled in
  hollow core optical fiber}.
\newblock \emph{Appl. Phys. Lett.} \textbf{2005}, 86~(12), 121106.

\bibitem{Phillips2014a}
Phillips, C.~L.; Jankowski, E.; Krishnatreya, B.~J.; Edmond, K.~V.; Sacanna,
  S.; Grier, D.~G.; Pine, D.~J.; Glotzer, S.~C.
\newblock {Digital colloids: reconfigurable clusters as high information
  density elements}.
\newblock \emph{Soft Matter} \textbf{2014}, 10~(38), 7468--7479.

\bibitem{Hadorn2010}
Hadorn, M.; {Eggenberger Hotz}, P.
\newblock {DNA-Mediated Self-Assembly of Artificial Vesicles}.
\newblock \emph{PLoS ONE} \textbf{2010}, 5~(3), e9886.

\bibitem{Wang2012}
Wang, Y.; Wang, Y.; Breed, D.~R.; Manoharan, V.~N.; Feng, L.; Hollingsworth,
  A.~D.; Weck, M.; Pine, D.~J.
\newblock {Colloids with valence and specific directional bonding}.
\newblock \emph{Nature} \textbf{2012}, 491~(7422), 51--55.

\bibitem{Wang2015a}
Wang, Y.; Wang, Y.; Zheng, X.; Ducrot, {\'{E}}.; Yodh, J.~S.; Weck, M.; Pine,
  D.~J.; Ducrot, E.; Yodh, J.~S.; Weck, M.; Pine, D.~J.
\newblock {Crystallization of DNA-coated colloids}.
\newblock \emph{Nature Communications} \textbf{2015}, 6, 7253.

\bibitem{Schade2013}
Schade, N.~B.; Holmes-Cerfon, M.~C.; Chen, E.~R.; Aronzon, D.; Collins, J.~W.;
  Fan, J.~A.; Capasso, F.; Manoharan, V.~N.
\newblock {Tetrahedral Colloidal Clusters from Random Parking of Bidisperse
  Spheres}.
\newblock \emph{Phys. Rev. Lett.} \textbf{2013}, 110~(14), 148303.

\bibitem{Zhang2017}
Zhang, Y.; McMullen, A.; Pontani, L.~L.; He, X.; Sha, R.; Seeman, N.~C.;
  Brujic, J.; Chaikin, P.~M.
\newblock {Sequential self-assembly of DNA functionalized droplets}.
\newblock \emph{Nat. Commun.} \textbf{2017}, 8~(1), 21.

\bibitem{Sugimoto1992}
Sugimoto, T.; Sakata, K.
\newblock {Preparation of monodisperse pseudocubic $\alpha$-Fe2O3 particles
  from condensed ferric hydroxide gel}.
\newblock \emph{J. Colloid Interface Sci.} \textbf{1992}, 152~(2), 587--590.

\bibitem{Rossi2011}
Rossi, L.; Sacanna, S.; Irvine, W. T.~M.; Chaikin, P.~M.; Pine, D.~J.;
  Philipse, A.~P.
\newblock {Cubic crystals from cubic colloids}.
\newblock \emph{Soft Matter} \textbf{2011}, 7~(9), 4139--4142.

\bibitem{Meester2016}
Meester, V.; Kraft, D.~J.
\newblock {Spherical, Dimpled, and Crumpled Hybrid Colloids with Tunable
  Surface Morphology}.
\newblock \emph{Langmuir} \textbf{2016}, 32~(41), 10668--10677.

\bibitem{VanDerWel2017TPM}
Van~der Wel, C.; Bhan, R.~K.; Verweij, R.~W.; Frijters, H.~C.; Gong, Z.;
  Hollingsworth, A.~D.; Sacanna, S.; Kraft, D.~J.
\newblock Preparation of Colloidal Organosilica Spheres through Spontaneous
  Emulsification.
\newblock \emph{Langmuir} \textbf{2017}, 33~(33), 8174--8180.

\bibitem{DohertyTBP}
Doherty, R.~P.; Kraft, D.~J.
\newblock One-pot surfactant-free synthesis of organosilica colloids with
  various surface functional groups. In preparation.

\bibitem{Appel2013}
Appel, J.; Akerboom, S.; Fokkink, R.~G.; Sprakel, J.
\newblock {Facile One-Step Synthesis of Monodisperse Micron-Sized Latex
  Particles with Highly Carboxylated Surfaces}.
\newblock \emph{Macromol. Rapid Commun.} \textbf{2013}, 34~(16), 1284--1288.

\bibitem{Cremer1999}
Cremer, P.~S.; Boxer, S.~G.
\newblock {Formation and Spreading of Lipid Bilayers on Planar Glass Supports}.
\newblock \emph{The J. Phys. Chem. B} \textbf{1999}, 103~(13), 2554--2559.

\bibitem{Wel2016}
Van~der Wel, C.; Vahid, A.; Saric, A.; Idema, T.; Heinrich, D.; Kraft, D.~J.
\newblock {Lipid membrane-mediated attraction between curvature inducing
  objects}.
\newblock \emph{Sci. Rep.} \textbf{2016}, 6, 32825.

\bibitem{Axelrod1976}
Axelrod, D.; Koppel, D.; Schlessinger, J.; Elson, E.; Webb, W.
\newblock {Mobility measurement by analysis of fluorescence photobleaching
  recovery kinetics}.
\newblock \emph{Biophys. J.} \textbf{1976}, 16~(9), 1055--1069.

\bibitem{trackpy}
Allan, D.~B.; Caswell, T.; Keim, N.~C.; Van~der Wel, C.~M.
\newblock trackpy: Trackpy v0.4.1, \textbf{2018}.

\bibitem{Wel2017}
Van~der Wel, C.~M.
\newblock Lipid Mediated Colloidal Interactions.
\newblock Casimir PhD Series, \textbf{2017}.

\bibitem{Sarfati2017}
Sarfati, R.; B{\l}awzdziewicz, J.; Dufresne, E.~R.
\newblock {Maximum likelihood estimations of force and mobility from single
  short Brownian trajectories}.
\newblock \emph{Soft Matter} \textbf{2017}, 13~(11), 2174--2180.

\bibitem{emcee}
{Foreman-Mackey}, D.; {Hogg}, D.~W.; {Lang}, D.; {Goodman}, J.
\newblock {emcee: The MCMC Hammer}.
\newblock \emph{ArXiv e-prints} \textbf{2013}, 125, 306.

\bibitem{Richter2006}
Richter, R.~P.; B{\'{e}}rat, R.; Brisson, A.~R.
\newblock {Formation of solid-supported lipid bilayers: an integrated view}.
\newblock \emph{Langmuir} \textbf{2006}, 22~(8), 3497--3505.

\bibitem{Sackmann1996}
Sackmann, E.
\newblock {Supported Membranes: Scientific and Practical Applications}.
\newblock \emph{Science} \textbf{1996}, 271~(5245), 43--48.

\bibitem{Raedler1995}
R{\"a}dler, J.; Strey, H.; Sackmann, E.
\newblock {Phenomenology and Kinetics of Lipid Bilayer Spreading on Hydrophilic
  Surfaces}.
\newblock \emph{Langmuir} \textbf{1995}, 11~(11), 4539--4548.

\bibitem{GoZen}
Gözen, I.; Jesorka, A.
\newblock Instrumental Methods to Characterize Molecular Phospholipid Films on
  Solid Supports.
\newblock \emph{Anal. Chem.} \textbf{2012}, 84~(2), 822--838.

\bibitem{Machan2010}
Mach{\'{a}}ň, R.; Hof, M.
\newblock {Lipid diffusion in planar membranes investigated by fluorescence
  correlation spectroscopy}.
\newblock \emph{Biochim. Biophys. Acta} \textbf{2010}, 1798~(7), 1377--1391.

\bibitem{Jing2014}
Jing, Y.; Trefna, H.; Persson, M.; Kasemo, B.; Svedhem, S.
\newblock {Formation of supported lipid bilayers on silica: relation to lipid
  phase transition temperature and liposome size.}
\newblock \emph{Soft Matter} \textbf{2014}, 10~(1), 187--95.

\bibitem{Castillo2014}
Castillo, S. I.~R.; Ouhajji, S.; Fokker, S.; Ern{\'{e}}, B.~H.; Schneijdenberg,
  C. T. W.~M.; Thies-Weesie, D. M.~E.; Philipse, A.~P.
\newblock {Silica cubes with tunable coating thickness and porosity: From
  hematite filled silica boxes to hollow silica bubbles}.
\newblock \emph{Microporous Mesoporous Mater.} \textbf{2014}, 195, 75--86.

\bibitem{Cha2006}
Cha, T.; Guo, A.; Zhu, X.-Y.
\newblock Formation of Supported Phospholipid Bilayers on Molecular Surfaces:
  Role of Surface Charge Density and Electrostatic Interaction.
\newblock \emph{Biophysical Journal} \textbf{2006}, 90~(4), 1270--1274.

\bibitem{DeGennes1987}
De~Gennes, P.~G.
\newblock {Polymers at an interface; a simplified view}.
\newblock \emph{Adv. Colloid Interface Sci.} \textbf{1987}, 27~(3-4), 189--209.

\bibitem{Upadhyayula2012a}
Upadhyayula, S.; Quinata, T.; Bishop, S.; Gupta, S.; Johnson, N.~R.; Bahmani,
  B.; Bozhilov, K.; Stubbs, J.; Jreij, P.; Nallagatla, P.; Vullev, V.~I.
\newblock {Coatings of polyethylene glycol for suppressing adhesion between
  solid microspheres and flat surfaces.}
\newblock \emph{Langmuir} \textbf{2012}, 28~(11), 5059--69.

\bibitem{VanDerWel2017}
Van~der Wel, C.; Bossert, N.; Mank, Q.~J.; Winter, M.~G.; Heinrich, D.; Kraft,
  D.~J.
\newblock Surfactant-free colloidal particles with specific binding affinity.
\newblock \emph{Langmuir} \textbf{2017}, 33~(38), 9803--9810.

\bibitem{VanDerWel2018}
{Van Der Wel}, C.; {Van De Stolpe}, G.~L.; Verweij, R.~W.; Kraft, D.~J.
\newblock {Micrometer-sized TPM emulsion droplets with surface-mobile binding
  groups}.
\newblock \emph{J. Phys.: Condensed Matter} \textbf{2018}, 30~(9).

\bibitem{naumann2002polymer}
Naumann, C.~A.; Prucker, O.; Lehmann, T.; R{\"u}he, J.; Knoll, W.; Frank, C.~W.
\newblock The polymer-supported phospholipid bilayer: Tethering as a new
  approach to substrate- membrane stabilization.
\newblock \emph{Biomacromolecules} \textbf{2002}, 3~(1), 27--35.

\bibitem{Tanaka2005}
Tanaka, M.; Sackmann, E.
\newblock {Polymer-supported membranes as models of the cell surface}.
\newblock \emph{Nature} \textbf{2005}, 437~(7059), 656--663.

\bibitem{Tanaka2006}
Tanaka, M.
\newblock {Polymer-Supported Membranes: Physical Models of Cell Surfaces}.
\newblock \emph{MRS Bulletin} \textbf{2006}, 31, 513--520.

\bibitem{Wagner2000}
Wagner, M.~L.; Tamm, L.~K.
\newblock {Tethered Polymer-Supported Planar Lipid Bilayers for Reconstitution
  of Integral Membrane Proteins: Silane-Polyethyleneglycol-Lipid as a Cushion
  and Covalent Linker}.
\newblock \emph{Biophys. J.} \textbf{2000}, 79~(3), 1400--1414.

\bibitem{Deverall2008}
Deverall, M.~A.; Garg, S.; Ludtke, K.; Jordan, R.; Ruhe, J.; Naumann, C.~A.
\newblock Transbilayer coupling of obstructed lipid diffusion in
  polymer-tethered phospholipid bilayers.
\newblock \emph{Soft Matter} \textbf{2008}, 4, 1899--1908.

\bibitem{lipowsky1995bending}
Lipowsky, R.
\newblock Bending of membranes by anchored polymers.
\newblock \emph{Europhys. Lett.)} \textbf{1995}, 30~(4), 197.

\bibitem{Geerts2010}
Geerts, N.; Eiser, E.
\newblock {DNA-functionalized colloids: Physical properties and applications}.
\newblock \emph{Soft Matter} \textbf{2010}, 6~(19), 4647--4660.

\bibitem{Valignat2005}
Valignat, M.-P.; Theodoly, O.; Crocker, J.~C.; Russel, W.~B.; Chaikin, P.~M.
\newblock {Reversible self-assembly and directed assembly of DNA-linked
  micrometer-sized colloids}.
\newblock \emph{Proc. Natl. Acad. Sci. U.S.A.} \textbf{2005}, 102~(12),
  4225--4229.

\bibitem{Biancaniello2007}
Biancaniello, P.~L.; Crocker, J.~C.; Hammer, D.~A.; Milam, V.~T.
\newblock {DNA-mediated phase behavior of microsphere suspensions}.
\newblock \emph{Langmuir} \textbf{2007}, 23~(5), 2688--2693.

\bibitem{Nordlund2009}
Nordlund, G.; Lönneborg, R.; Brzezinski, P.
\newblock {Formation of Supported Lipid Bilayers on Silica Particles Studied
  Using Flow Cytometry}.
\newblock \emph{Langmuir} \textbf{2009}, 25~(8), 4601--4606.

\bibitem{Tan2002a}
Tan, A.; Ziegler, A.; Steinbauer, B.; Seelig, J.
\newblock {Thermodynamics of sodium dodecyl sulfate partitioning into lipid
  membranes}.
\newblock \emph{Biophys. J.} \textbf{2002}, 83~(3), 1547--1556.

\bibitem{TanjaDrobek2005}
Drobek, T.; Spencer, N.~D.; Heuberger, M.
\newblock {Compressing PEG brushes}.
\newblock \emph{Macromolecules} \textbf{2005}, 38~(12), 5254--5259.

\bibitem{Meng2004}
Meng, F.; Engbers, G. H.~M.; Feijen, J.
\newblock {Polyethylene glycol-grafted polystyrene particles.}
\newblock \emph{J. Biomed. Mater. Res. Part A} \textbf{2004}, 70~(1), 49--58.

\bibitem{Garbuzenko2005}
Garbuzenko, O.; Barenholz, Y.; Priev, A.
\newblock {Effect of grafted PEG on liposome size and on compressibility and
  packing of lipid bilayer}.
\newblock \emph{Chem. Phys. Lipids} \textbf{2005}, 135~(2), 117--129.

\bibitem{Angioletti-Uberti2014}
Angioletti-Uberti, S.; Varilly, P.; Mognetti, B.~M.; Frenkel, D.
\newblock {Mobile linkers on DNA-coated colloids: Valency without patches}.
\newblock \emph{Phys. Rev. Lett.} \textbf{2014}, 113~(12), 128303.

\bibitem{Lipowsky2013}
Lipowsky, R.
\newblock {Spontaneous tubulation of membranes and vesicles reveals membrane
  tension generated by spontaneous curvature}.
\newblock \emph{Faraday Discuss.} \textbf{2013}, 161, 305--331.

\bibitem{Nakano1999}
Nakano, S.~I.; Fujimoto, M.; Hara, H.; Sugimoto, N.
\newblock {Nucleic acid duplex stability: Influence of base composition on
  cation effects}.
\newblock \emph{Nucleic Acids Res.} \textbf{1999}, 27~(14), 2957--2965.

\end{thebibliography}


\begin{thebibliography}{1}
\expandafter\ifx\csname url\endcsname\relax
  \def\url#1{\texttt{#1}}\fi
\expandafter\ifx\csname urlprefix\endcsname\relax\def\urlprefix{URL }\fi
\providecommand{\bibinfo}[2]{#2}
\providecommand{\eprint}[2][]{\url{#2}}

\bibitem{Hunter:2007}
\bibinfo{author}{Hunter, J.~D.}
\newblock \bibinfo{title}{Matplotlib: A 2d graphics environment}.
\newblock \emph{\bibinfo{journal}{Comput. Sci. Eng.}}
  \textbf{\bibinfo{volume}{9}}, \bibinfo{pages}{90--95} (\bibinfo{year}{2007}).

\bibitem{Chakraborty2016a}
\bibinfo{author}{Chakraborty, I.}, \bibinfo{author}{Meester, V.},
  \bibinfo{author}{Van~der Wel, C.} \& \bibinfo{author}{Kraft, D.~J.}
\newblock \bibinfo{title}{{Colloidal joints with designed motion range and
  tunable joint flexibility}}.
\newblock \emph{\bibinfo{journal}{Nanoscale}} \textbf{\bibinfo{volume}{9}},
  \bibinfo{pages}{7814--7821} (\bibinfo{year}{2017}).

\end{thebibliography}

\onecolumn
\begin{figure}

    \centering
    \includegraphics[width=3.25in]{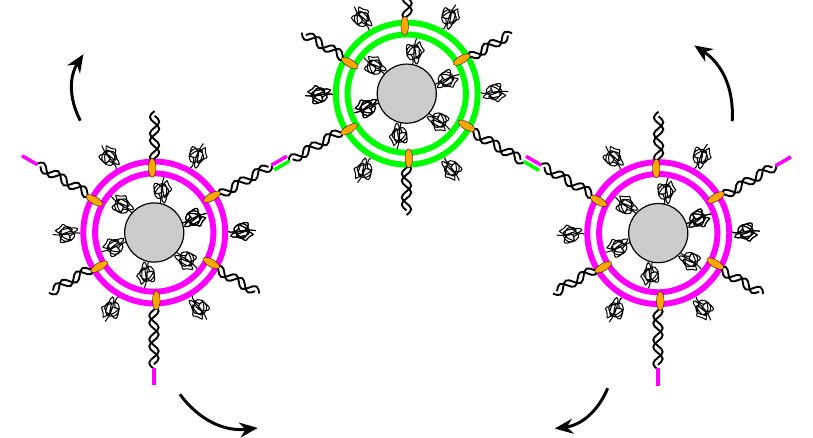} \caption{For the Table of Contents only.}

\end{figure}

\end{document}


\maketitle

\begingroup
    \let\clearpage\relax
    \let\cleardoublepage\relax

    \listoffigures
    \listoftables

\endgroup

\appendix

\begin{figure}

    \includegraphics[width=\linewidth]{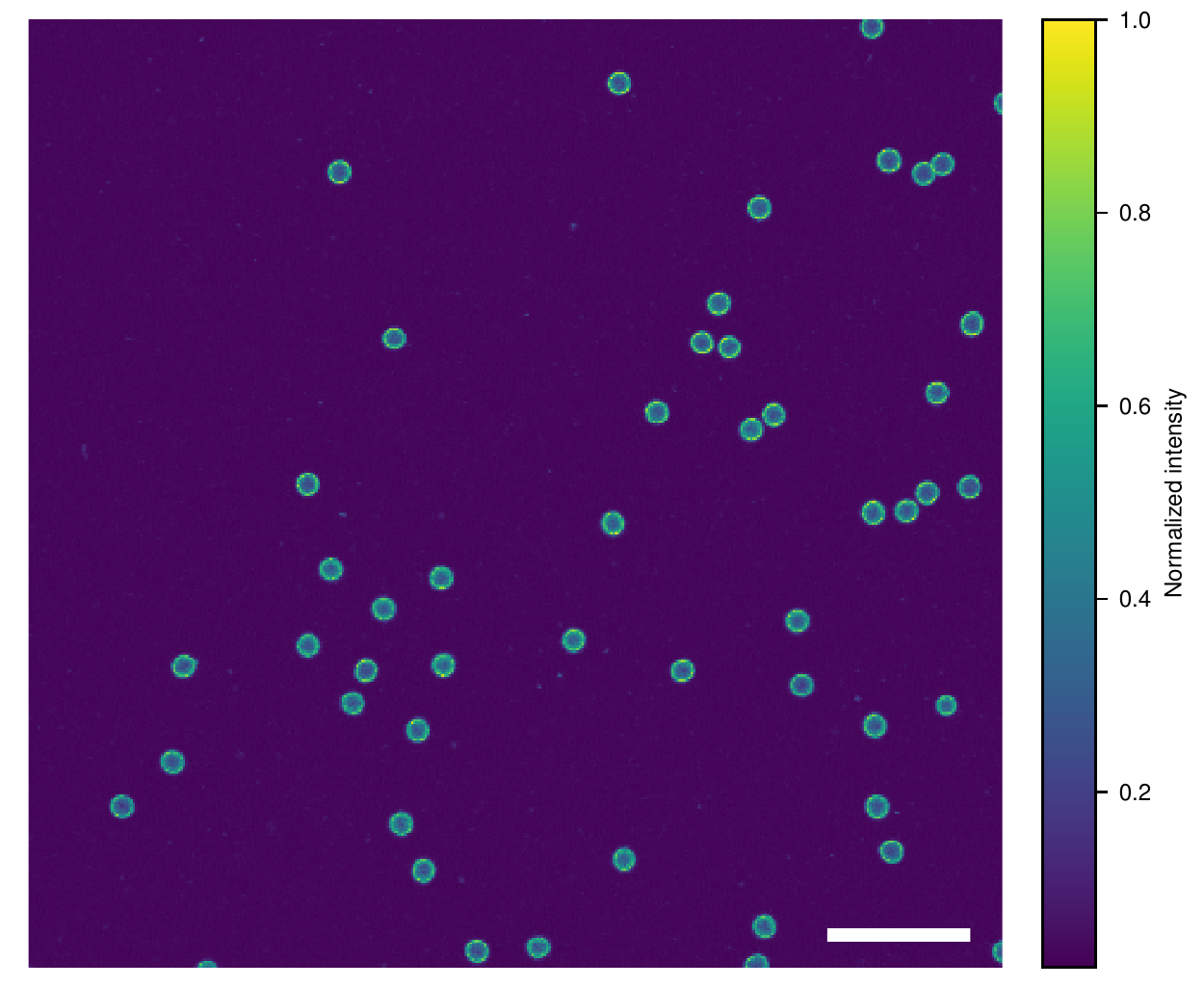}

    \caption[Overview picture of CSLBs]{Overview picture showing the
    homogeneity of the lipid bilayer for CSLBs containing \SI{1}{\mole\percent}
DOPE-PEG(2000). The scale bar is \SI{15}{\um}.\label{fig:overview}}

\end{figure}

\begin{table}

    \centering
    \caption[List of DNA strands]{Summary of all DNA strand sequences and their names. Sticky ends
    are marked in cursive. \label{tab:dna}}\vspace{1em}

    \begin{tabularx}{\textwidth}{llX}
        \toprule
        \textbf{No.} & \textbf{Name} & \textbf{Sequence} \\ \midrule
        1 & \SI{10}{\nm} Base & \texttt{Cholesterol\--TEG\--3$'$\--GTT\--AGC\--CCG\--ATT\--ACA\--GAG\--CGT\--TCT\--TT\--3$'$}\\
        2 & \SI{10}{\nm} Inert & \texttt{Cholesterol\--TEG\--5$'$\--TTT\--GAA\--CGC\--TCT\--GTA\--ATC\--GGG\--CTA\--AC\--3$'$}\\
        3 & \SI{20}{\nm} Base & \texttt{Cholesterol\--TEG\--3$'$\--TTT\--TAG\--CGA\--TGG\--GAA\--GCG\--TGT\--CAG\--TTA\--GAT\--CTC\--TCG\--GGA\--CGG\--AAT\--GC\--5$'$}\\
        4 & \SI{20}{\nm} Inert & \texttt{Cholesterol\--TEG\--5$'$\--TTT\--ATC\--GCT\--ACC\--CTT\--CGC\--ACA\--GTC\--AAT\--CTA\--GAG\--AGC\--CCT\--GCC\--TTA\--CGA\-- 3$'$} \\
        5 & \SI{20}{\nm} Single Linker A & \texttt{Cholesterol\--TEG\--5$'$\--TTT\--ATC\--GCT\--ACC\--CTT\--CGC\--ACA\--GTC\--AAT\--CTA\--GAG\--AGC\--CCT\--GCC\--TTA\--CGA\--\textit{GTA\--GAA\--GTA\--GG}\--3$'$\--6FAM}\\
        6 & \SI{20}{\nm} Single Linker A$'$ & \texttt{Cholesterol\--TEG\--5$'$\--TTT\--ATC\--GCT\--ACC\--CTT\--CGC\--ACA\--GTC\--AAT\--CTA\--GAG\--AGC\--CCT\--GCC\--TTA\--CGA\--\textit{CCT\--ACT\--TCT\--AC}\--3$'$\--Cy3}\\
        7 & \SI{30}{\nm} Base & \texttt{5$'$\--TCG\--TAA\--GGC\--AGG\--GCT\--CTC\--TAG\--ACA\--GGG\--CTC\--TCT\--GAA\--TGT\--GAC\--TGT\--GCG\--AAG\--GTG\--ACT\--GTG\--CGA\--AGG\--GTA\--GCG\--ATT\--TT\--3$'$}\\
        8 & \SI{30}{\nm} Single Linker A & \texttt{Double Stearyl\--HEG\--5$'$\--TT\--TAT\--CGC\--TAC\--CCT\--TCG\--CAC\--AGT\--CAC\--CTT\--CGC\--ACA\--GTC\--ACA\--TTC\--AGA\--GAG\--CCC\--TGT\--CTA\--GAG\--AGC\--CCT\--GCC\--TTA\--CGA\--\textit{GTA\--GAA\--GTA\--GG}\--3$'$\--6FAM}\\
        9 & \SI{30}{\nm} Single Linker A$'$ & \texttt{Double Stearyl\--HEG\--5$'$\--TT\--TAT\--CGC\--TAC\--CCT\--TCG\--CAC\--AGT\--CAC\--CTT\--CGC\--ACA\--GTC\--ACA\--TTC\--AGA\--GAG\--CCC\--TGT\--CTA\--GAG\--AGC\--CCT\--GCC\--TTA\--CGA\--\textit{CCT\--ACT\--TCT\--AC}\--3$'$\--Cy3} \\
        \bottomrule
    \end{tabularx}
\end{table}

\begin{figure}

    \includegraphics[width=\linewidth]{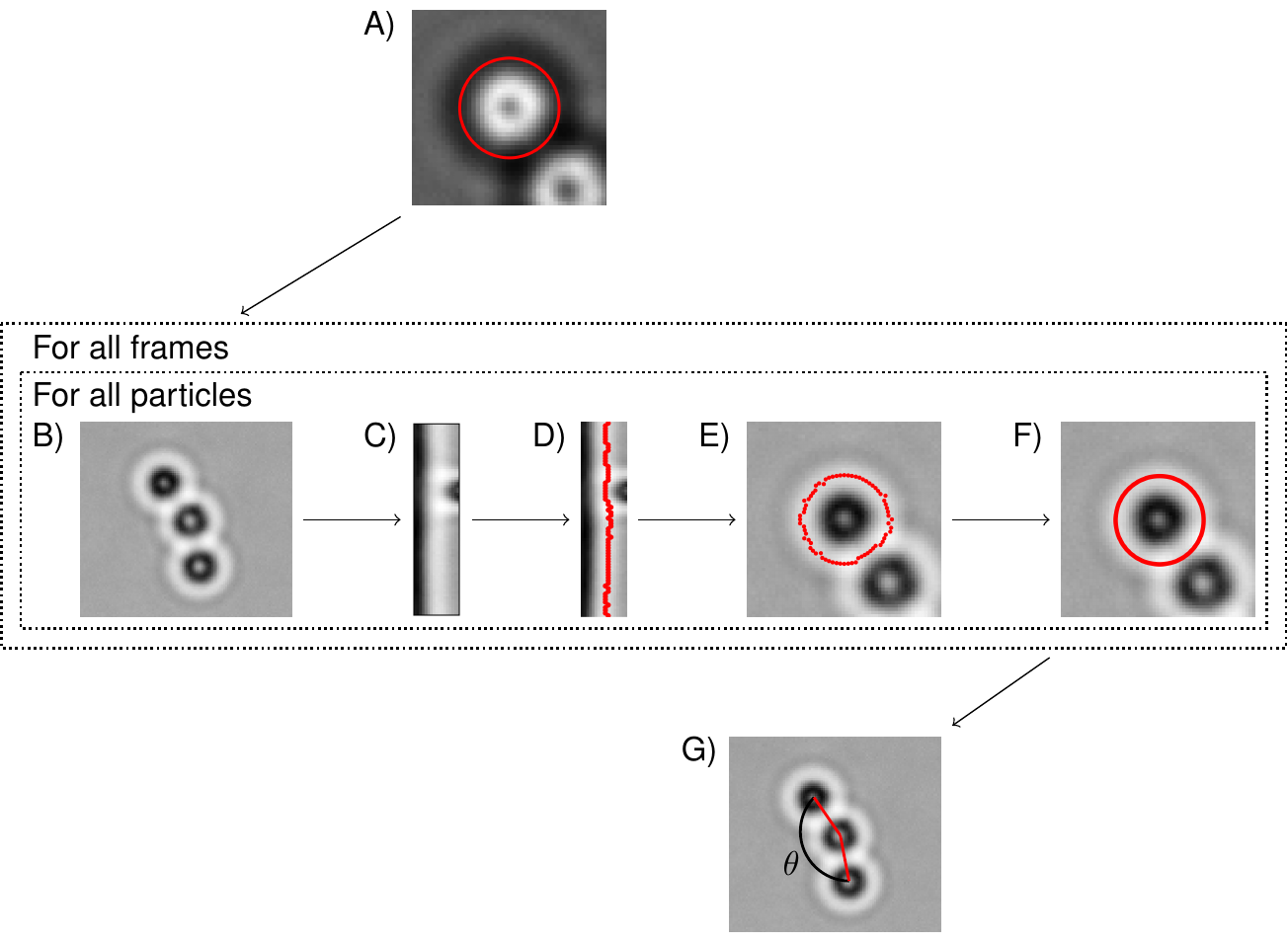}

    \caption[Schematic depiction of tracking algorithm]{The algorithm used for tracking particles in bright-field movies
        depicted graphically. \textbf{A)} The user is asked to manually select
        the particles that need to be tracked from the first frame using a
        Matplotlib \cite{Hunter:2007} interface. \textbf{B)} The current frame
        is inverted so that the dark ring around features becomes bright.
        \textbf{C)} The frame is interpolated and converted to polar
        coordinates with the current provisional particle position at the
        origin. \textbf{D)} For each row (corresponds to each polar angle), the
        position with the maximum intensity is found (for intensities higher
        than a set threshold). \textbf{E)} The coordinates that were found are
        then converted to the original Cartesian coordinates. \textbf{F)} A
        circle is fit to the coordinates using a least squares method.
        \textbf{G)} The opening angle between the three particles is determined
    using simple trigonometry, whilst keeping the particle order the same for
all frames.  \label{fig:particle_tracking}}

\end{figure}

\begin{figure}

    \centering
    \includegraphics[scale=1]{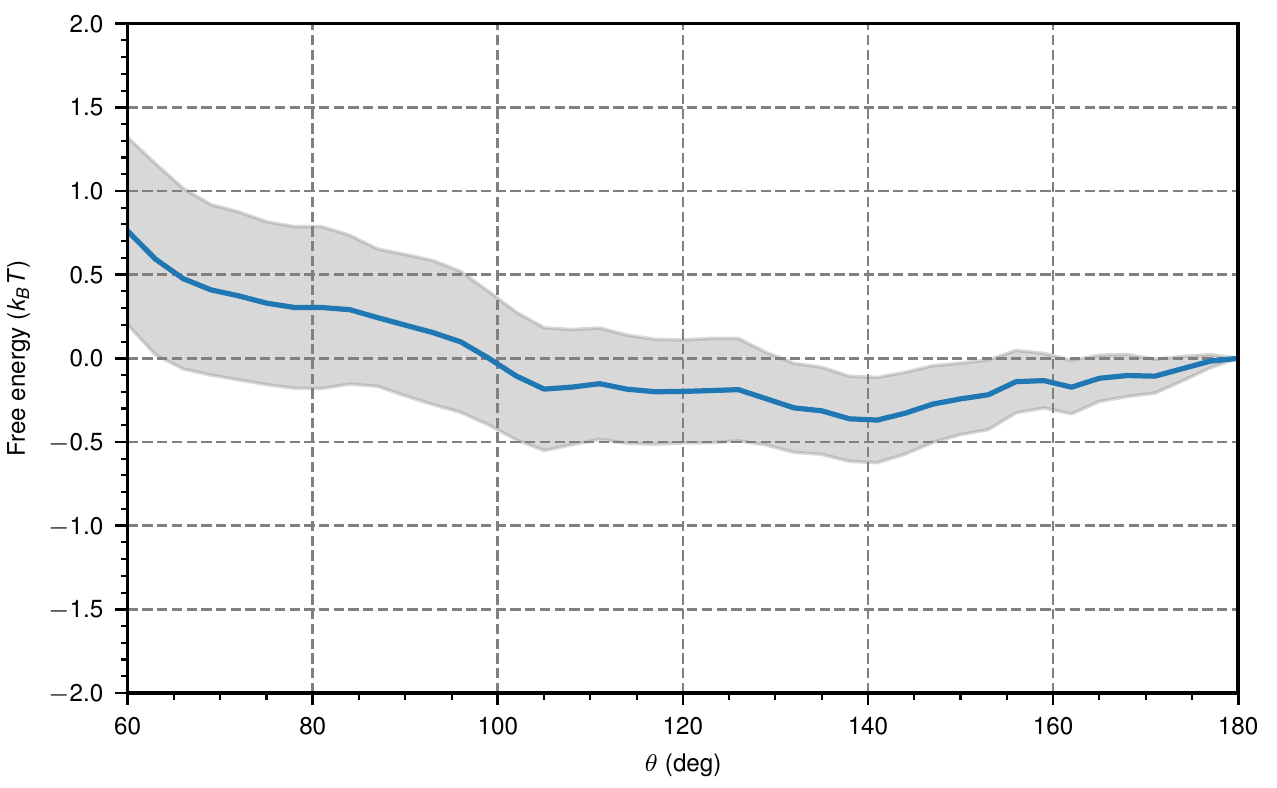}

    \caption[Previous results re-analyzed using a maximum likelihood method]{In \cite{Chakraborty2016a}, we measured a preferred angle of
    \SI{140}{\degree} with a magnitude of roughly 0.9 $K_B T$. When we analyze
this data using the angular displacement method outlined in this work, we see
that the observed preference is within the experimental error (indicated by the shaded gray area).
\label{fig:indrani}}

\end{figure}

\begin{figure}

    \includegraphics[width=\linewidth]{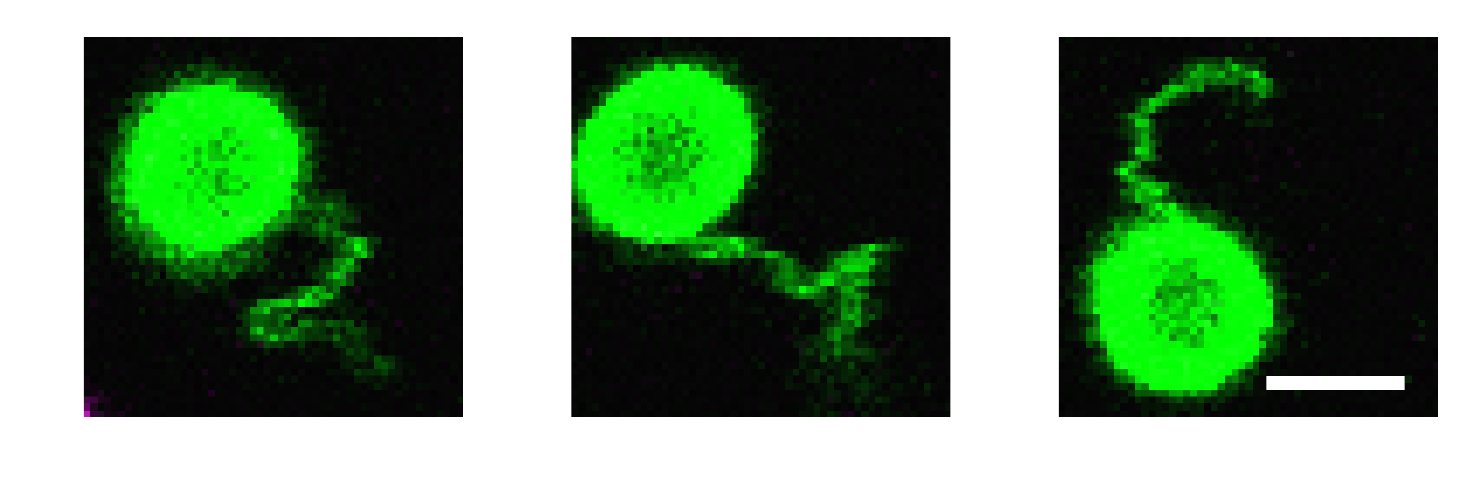}

    \caption[CSLBs with membrane tubes]{Figure showing the membrane tubes that form at high DNA coating
        concentrations for three different particles. The scale bar is
        \SI{2}{\um}. The tubes are comparable to the particle size
        (\SI{2}{\um}) and are very floppy. Note that the brightness was
    increased to show the tubes and as a result the particles are
oversaturated.\label{fig:tubes}}

\end{figure}

\clearpage
\bibliography{references}{}